\documentclass[final]{jfm}

\usepackage{color,soul}
\usepackage[dvipsnames,xcdraw]{xcolor}
\usepackage{graphicx}
\usepackage{newtxtext}
\usepackage{newtxmath}
\usepackage{natbib}
\usepackage[hidelinks]{hyperref}
\usepackage[noabbrev,nameinlink]{cleveref}
\usepackage{bm}% bold math
\usepackage{amsfonts,amsmath} 
\usepackage{textcomp}
\usepackage{epsfig}
\usepackage{verbatim}
\usepackage{physics}
\usepackage{siunitx}
\usepackage{algorithm}
\usepackage{algpseudocode}
\usepackage{subcaption}
\usepackage[normalem]{ulem}
\usepackage{mathtools,caption}
\usepackage{afterpage}
\usepackage{circledsteps}
\usepackage{pgfplots}\pgfplotsset{compat=1.17}%1.18
\usepackage{dcolumn}% Align table columns on decimal point
\usepackage{float}
\usepackage{hhline}
\newcolumntype{P}[1]{>{\centering\arraybackslash}p{#1}}
\newcommand{\RNum}[1]{\uppercase\expandafter{\romannumeral #1\relax}}
\crefformat{section}{\S#2#1#3} % see manual of cleveref, section 8.2.1
\crefformat{subsection}{\S#2#1#3}
\crefformat{subsubsection}{\S#2#1#3}

\newcommand{\RomanNumeralCaps}[1]
\linenumbers

\title{Smart Skin separation control using distributed-input distributed-output, multi-modal actuators, and machine learning}

 \author{Songqi Li\aff{1}
 \corresp{\email{lisongqi@hit.edu.cn}}}

\affiliation{\aff{1}School of Mechanical Engineering and Automation, Harbin Institute of Technology, 518055 Shenzhen, PR China
}

\begin{document}
\maketitle

\begin{abstract}
Efficient flow separation control represents significant economic benefit.
This study applies a machine learning algorithm to minimize flow separation in Smart Skin, a flow control device that features distributed-input and distributed-output (DIDO). 
Smart Skin comprises 30 hybrid actuator units, each integrating a height-adjustable vortex generator and a mini-jet actuator. 
These units are deployed on a backward-facing ramp to reduce flow separation in a distributed manner. 
To monitor the flow state, distributed pressure taps are deployed around the multi-modal actuators. 
Parametric studies indicate that the mapping between control parameters and separation control performance is complex. 
To optimize separation control, a cutting-edge variant of the particle swarm optimization (PSO-TPME, \citealt{Shaqarin2023}) are used for the control parameters in the Smart Skin. 
This algorithms is capable of achieving a fast optimization in high-dimensional parameter spaces. 
The results demonstrate the efficiency of PSO-TPME, and the optimized solution significantly outperform the best result from the parametric study. 
These findings represent a promising future of machine learning-based flow control using distributed actuators and sensors.

\end{abstract}

\begin{keywords}
Flow control; distributed-input distributed-output; multi-modal actuator; machine learning
\end{keywords}

%{\bf MSC Codes }  {\it(Optional)} Please enter your MSC Codes here

\section{Introduction}
\label{sec:intro}

Turbulence control offers substantial economic and environmental benefits in the realm of transportation vehicles.
In the case of high-speed trains,
The aerodynamic drag takes up 75-80 \% of the total drag when the train is running at a speed of \SI{300}{\kilo\meter\per\hour} \citep{Baker2014,Raina2017,Zhang2018}. 
This proportion increases to over 90 \% when the train speed reaches \SI{400}{\kilo\meter\per\hour}  \citep{Yang2012,Yu2021}. 
In passenger cars, aerodynamic pressure drag is the predominant factor \citep{Sudin2014,Altaf2014,Geropp2000}.
The contribution of aerodynamic pressure drag corresponds to 90 \% of the total aerodynamic drag, with 80 \% of this contribution originating from the rear part of the car \citep{Kourta2012}.
One of the main sources of aerodynamic drag in passenger cars is the flow separation near the vehicle's rear end, where the detached flow often results in significant energy losses \citep{Hucho1993}.

A wide range of flow control techniques have been successfully employed in both academic and industrial flow configurations to mitigate flow separation \citep{Brunton2015,GadelHak1996}.
Passive control devices, including vortex generators \citep{Selby1992,Lin1994}, dimples \citep{Lake2000,Ballerstein2023}, flaps \citep{CamachoSnchez2023}, and transverse grooves \citep{Mariotti2017}, have demonstrated their effectiveness for separation control in various flow configurations.
In a recent study by \citet{Viswanathan2021}, a parametric analysis was conducted to explore the effect of different vortex generators on car drag reduction under varying yawing angles.
In recent decades, the rapid advancement of actuators has made active flow control a prominent research area. 
A comprehensive overview of active actuators can be found in \citet{Cattafesta2011}. 
Fluidic oscillators \citep{Raghu2013,Metka2015}, synthetic jets \citep{Gilarranz2005,Kim2009}, plasma actuators \citep{Post2006,Roupassov2009}, pulsed jets \citep{Li2017,Fan2020}, have been successfully implemented in various flow configurations, yielding significant improvements in separation control.
Despite the additional energy investment required, the promising performance, flexibility, and robustness of active control methods have made them a popular choice for flow control applications \citep{ScottCollis2004}.

Synergizing both active and passive flow control devices represents a promising future for more effective flow control outcomes.
In \citet{Yagiz2012}, a 5 \% aerodynamic drag reduction was achieved in a 2D airfoil by combining bump and jet actuation.
This combined approach yields significantly better results compared to using either form of flow control strategy alone. 
Similarly, \citet{Liesner2014} simultaneously applied boundary layer fences and suction actuation in a compressor cascade to reduce energy loss, demonstrating the effectiveness of combining both flow control strategies.
However, in these pioneering studies, passive and active control devices were deployed independently and lacked effective integration. 
The potential for achieving even more promising flow control outcomes with minimal energy investment lies in the development of multi-modal actuators, which are capable of performing active control based on passive control results. 
By integrating both active and passive elements, these multi-modal actuators have the potential to perform more effective flow control.

In the past decade, sensor technology has experienced rapid development. 
The trend of low cost, small size, and high reliability of sensors has led to an increasing focus on closed-loop flow control methods that utilize sensor signals for feedback \cite{Pinier2007, Pastoor2008, Staats2016}. 
Compared to open-loop control strategies, sensor feedback, as well as the represented flow stat, can further enhance flow control performance and effectively reduce energy investment \citep{Brunton2015}.
The simplest form of closed-loop control is the single-input single-output (SISO) system \citep{Sipp2010,Nair2021}. 
In recent years, there has been a surge in research focusing on flow control with multi-input, multi-output (MIMO) systems on the order of $\mathcal{O}(10)$. 
Numerous studies have investigated MIMO flow control, employing a range of model-based and model-free control algorithms in both experimental and numerical setups \citep{Samimy2007,luchtenburg2010,Aleksic2010,Sasaki2017,Toedtli2019,FanDW2020,Becker2007,Li2017,Zhou2020,CornejoMaceda2021}, among many others.

The inherent complexity of industrial flow configurations indeed calls for the expansion of sensor and actuator dimensions. 
To effectively control turbulent flows, the utilization of a distributed-input distributed-output (DIDO) system becomes crucial. 
This involves the deployment of a large number of actuators and sensors throughout the flow control plant.
Pioneering studies have exemplified the effectiveness of distributed flow control approaches. 
For instance, in the field of wind energy, vortex generators have been massively deployed on a wind turbine blade in a distributed manner to improve the aerodynamic performance \citep{MuellerVahl2012}. 
Similarly, for the active flow control of the bluff body wake, distributed minijet actuators have been employed to manipulate the wake structure and reduce drag \citep{Zigunov2022}. 
As the number of sensors and actuators expands in DIDO configurations, the development of dedicated flow control algorithms becomes increasingly crucial for efficient flow control. 

The ongoing machine learning revolution is transforming the field of flow control. 
Machine learning-based approaches have emerged as powerful tools for handling complex, non-linear, and non-convex flow control problems. 
These algorithms leverage the capabilities of machine learning algorithms to efficiently explore and exploit the parameter and functional spaces, identifying optimal control strategies within limited time horizons.
One of the key advantages of machine learning-based flow control algorithms is their ability to operate without a pre-existing model. 
Instead, they rely solely on input-output data gathered from the flow  control plant. 
These data-driven approaches can adapt to the inherent complexity and non-convexity of the plants, making them particularly well-suited for flow control applications.
For closed-loop control, two prevalent model-free techniques are machine learning control (MLC,  \citealt{Duriez2017,CornejoMaceda2021,xmlc2022}) and reinforcement learning (RL, \citealt{Verma2018,Rabault2019,Fan2020,Vinuesa2022}). 
In terms of open-loop control, machine learning optimization algorithms have been widely employed to obtain optimal flow control parameters across different applications \citep{Koumoutsakos2001,Li2020,Li2021,Zigunov2022,Paschereit2003}.

In this study, we present the design and development of the Smart Skin, an experimental separation control plant with distributed multi-modal actuators and distributed sensors.
These multi-modal actuators can simultaneously execute passive and active flow control, while the distributed sensors enable real-time monitoring of the flow state.
By controlling the on-off state of the multi-modal actuators, we conducted a parametric study where the flow control performances were examined under various arrangements. 
As a key innovation, we incorporated a machine learning algorithm to enhance the flow control outcomes beyond the parametric results. 
In particular, we employed a state-of-the-art variant of the particle swarm optimization (PSO-TPME \citealt{Shaqarin2023}), and achieve an additional 60 \% improvement in pressure recovery compared to the simple actuation pattern from a parametric study.
These encouraging results shed light on the future development of flow control experiments with distributed inputs and distributed outputs, leveraging the advantages of cutting-edge flow control hardware and machine learning algorithms.

This manuscript is organized as follows. We present the Smart Skin separation control experiment in \S~\ref{sec:S2}, including the hardware configuration in \S~\ref{ssec:S2.1}, the optimization problem and the optimizer in \S~\ref{ssec:S2.2}, and other relevant experimental instrumentation in \S~\ref{ssec:S2.3}.
The characterization of the Smart Skin, including the baseline flow profiles and a parametric study of some simple actuation patterns, are presented in \Cref{ssec:S3.1} and \Cref{ssec:S3.2}, respectively.
In \Cref{ssec:S3.3}, we present the results of flow control optimization, including the optimized control parameters and the corresponding flow field.
We close the manuscript by the conclusions and outlook in \Cref{sec:S4}.
\section{Methodology}\label{sec:S2}
\subsection{Smart Skin separation control using distributed-input distributed-output and multi-modal actuators}\label{ssec:S2.1}

%%%%%%%%%%%%%BEGIN %FIGURE%%%%%%%%%%%%%%%%
\begin{figure}
\centering
\begin{subfigure}[b]{0.5\linewidth}
\centering
\includegraphics[width=.99\linewidth]{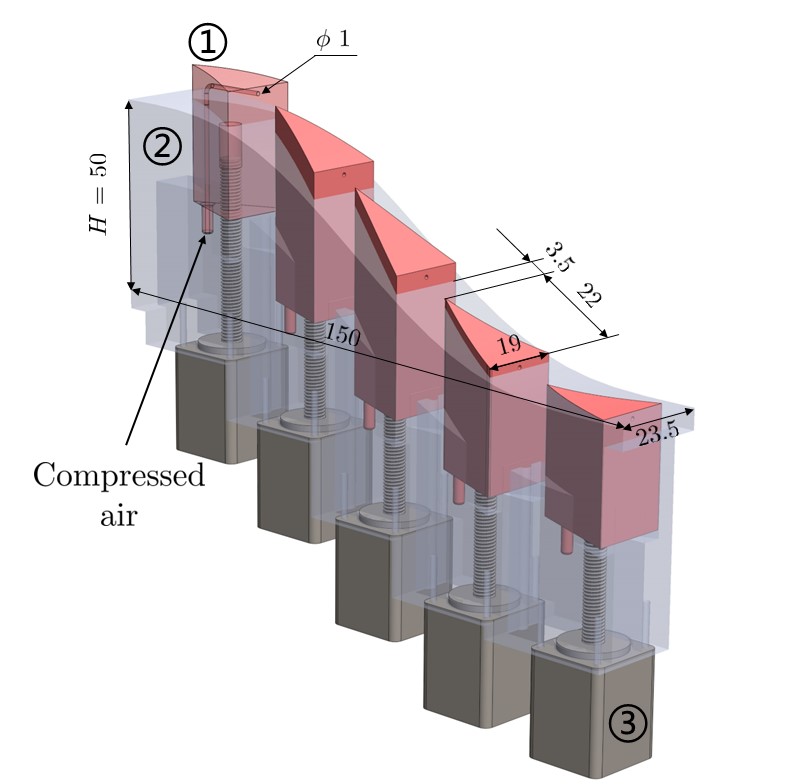}
\caption{}
\label{fig:SS1}
\end{subfigure}\\
\begin{subfigure}[b]{0.8\linewidth}
\centering
\includegraphics[width=.99\linewidth]{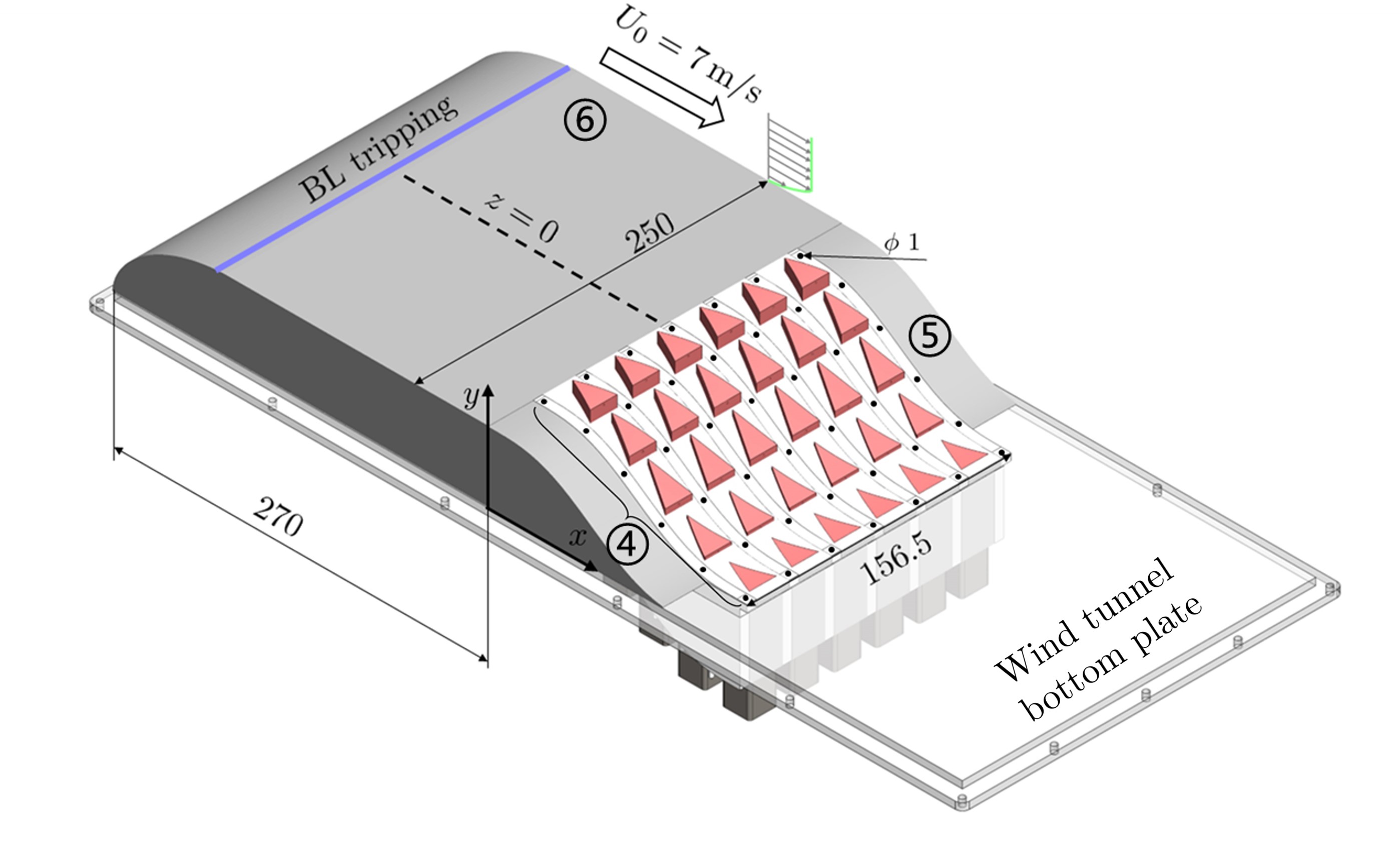}
\caption{}
\label{fig:SS2}
\end{subfigure}
\caption{The Smart Skin separation control device examplified for a smooth ramp. Key components are listed from \Circled{1} to \Circled{6}.
\Circled{1}: a multi-modal actuator acting as a vortex generator and a minijet injector; 
\Circled{2}: a ramp-shaped base installing the multi-modal actuators; 
\Circled{3}: a stepper motor controlling the height of the multi-modal actuator above;
\Circled{4}: pressure taps for static pressure measurements on the ramp surface;
\Circled{5}: the fixed ramp profile on the side; 
\Circled{6}: the front body with the boundary layer tripping device.
Key design parameters are also provided for reference (unit:\SI{}{\milli\meter}).
}
\label{fig:SS}
\end{figure}
%%%%%%%%%%%%%%%END %FIGURE%%%%%%%%%%%%%%%%

One key innovation in this study is the design and development of the Smart Skin, a separation control device that incorporates distributed-input, distributed-output, and multi-modal actuators. 
Our approach involves deploying numerous actuators and sensors across the flow separation surface in a distributed manner. 
In the Smart Skin configuration, each actuator is capable to mitigate flow separation at its respective location, while the distributed sensors provide local and global monitoring of the flow state.
Meanwhile, we propose and implement multi-modal actuators in the Smart Skin. 
These actuators possess the ability to execute both active and passive flow control commands. 
In scenarios where flow separation is not present, these actuators can be commanded to remain flush with the profiled wall. 
When local flow separation occurs, these actuators can act as passive devices and mitigate flow separation without additional energy investment.
Building upon the effectiveness of passive control, the implementation of additional active control techniques can be superimposed to harness the combined benefits of both flow control strategies. 

Based on the above-mentioned design principles, we implement the Smart Skin based on a backward-facing ramp model.
This classical model has been intensively studied in \citet{Jukes2011,Lardeau2011,Debien2016,Kourta2015,Dandois2007}, among many others. 
Reducing flow separation and recirculation in this simplified model has strong implications to industrial flow applications.
The current research considers a rounded backward-facing step profile which was initially proposed in \citet{Dandois2007}.
The mathematical expression of the ramp profile is given as follows:
\begin{equation}
    \frac{y}{H} = \frac{1}{2\pi}\biggl(\text{sin}\biggl( \frac{a\pi x}{H} \biggr) -\frac{a\pi x}{H} \biggr)+1, \ \ \frac{x}{H}\in \biggl[0, \frac{2}{a} \biggr]. 
    \label{eqn:eq1}
\end{equation}
In this equation, $x$ and $y$ represent streamwise and wall-normal directions, respectively. $H$ is the step height and is \SI{50}{\milli\meter} in the current implementation.
The shape factor $a=0.703$ is adopted such that the maximum slope is \SI{35}{\degree}.

The design of the Smart Skin is based on the ramp profile described in \Cref{eqn:eq1}. \Cref{fig:SS} presents the three-dimensional schematics of the Smart Skin.
The Smart Skin is composed of massively deployed multi-modal actuators and sensors on the ramp surface.
The design follows a modular approach, as depicted in Figure \ref{fig:SS1}. 
Each module comprises five equidistantly arranged multi-modal actuators that conform to the ramp profile. 
To achieve complex inner and outer geometries, these actuators are fabricated using 3D-printing techniques. 
Each multi-modal actuator takes the shape of a triangular prism, with its top surface following the local ramp profile.
At the bottom of the each multi-modal actuator, a stepper motor is installed to move the actuator in the wall-normal direction.
When the actuator's top surface coincides with the ramp profile, it remains inactive.
However, the actuator can act as a passive vortex generator when the actuator is lifted up.
These triangular prisms are capable to produce a pair of counter-rotating vortices from the sharp leading edge, and the effectiveness of similar configurations has been validated in \citet{Tian2020, Viswanathan2021}. 
The motion of the stepper motor is quasi-static, and different passive actuation patterns can be created by assigning different heights to the actuators.
The actuator dimensions are designed to maximize the number of actuators accommodated on the ramp body. 
The height of each actuator can be adjusted between 0 and \SI{10}{\milli\meter}. 
A height of 0 signifies that the actuator is inactive with the ramp surface, while a maximum height of \SI{10}{\milli\meter} surpasses the boundary layer thickness ($\delta_{.99}\approx\SI{9.5}{\milli\meter}$) of the incoming flow.

To achieve active control, each actuator is embedded with an interior air tubing. 
The tubing has a diameter of \SI{1}{\milli\meter}, and its outlet is positioned \SI{2}{\milli\meter} below the actuator's top surface. 
We connect the tubing inlets of all actuators to a plenum chamber that is pressurized with a compressed air supply. 
This configuration allows the actuators to issue high-momentum air in the downstream direction, functioning similarly to minijet injectors.
To regulate the supply pressure of all actuators, we utilize an SMC-ITV2050 pressure regulator. 
We careful adjust the supply pressure and ensure a consistent volumetric flow rate of \SI{60}{\liter\per\minute} when all active actuators are activated.
To control the on-off state of each actuator in a timely manner, we also incorporate high-frequency solenoid valves (Festo, MHE2-MS1H) between each actuator and the compressed air supply. 
These solenoid valves have the capability to operate at frequencies ranging from 0 to \SI{200}{\hertz}, enabling steady blowing, periodic blowing, and even user-defined actuation patterns within the system.

As displayed in \Cref{fig:SS2}, the distributed deployment of multi-modal actuators is achieved by stacking a total of six modules in the spanwise direction. This configuration results in a total of $N=30$ equally-spaced multi-modal actuators on the ramp surface.
In addition to the actuators, we also incorporate a staggered distribution of 6 (streamwise) $\times$ 7 (spanwise) pressure taps among the multi-modal actuators. 
Each pressure tap is positioned at the center of a neighborhood of actuators, and has a diameter of \SI{1}{\milli\meter}. 
To monitor the pressure distribution on the ramp surface,
all pressure taps are connected to a HangHua PSU pressure scanner using plastic tubing of \SI{30}{\centi\meter} long with an innder diameter of \SI{2}{\milli\meter}.
With this setup, we can execute passive, active, and hybrid control commands while simultaneously evaluating the wall pressure distribution under the corresponding control commands.

The assembled actuators and sensors are installed in the test section of an open-loop wind tunnel. 
To ensure a uniform incoming flow, a front body with a streamlined nose is installed ahead of the assembly to ensure a uniform incoming flow.
In addition, boundary layer tripping device is installed in the upstream location to trigger a turbulent boundary layer in the incoming flow.
To avoid the region of strong wall effect, the multi-modal actuators are not deployed near the side walls of the wind tunnel. 
Instead, solid parts with the identical ramp profile are used to fill the corresponding regions. 
For this particular study, the freestream velocity is set to \SI{7}{\meter\per\second}. This velocity corresponds to a Reynolds number of $Re_H=23,000$, based on the height of the ramp. 

A control system based on LabVIEW is developed to trigger the actions of the multi-modal actuators according to the user-defined commands. 
The control system is implemented in a PXIe-8840 Real-time system, enabling quasi-static control of the actuator heights, as well as open- and closed-loop control of the active actuations.
To control the heights of the multi-modal actuators, as well as the on-off states of the active excitations, an NI PXI-6509 digital I/O module is employed. 
This module provides the necessary digital ports to send control signals to the passive and active devices, enabling precise actuation of the multi-modal actuators.
For data acquisition, an NI PXI-6225 module is utilized. 
This module collects voltage signals obtained from the pressure scanner. 
These voltage signals can be converted into pressure readings ($\bm{s}$) through a careful calibration process. 
In the subsequent discussions, the height of each actuator is denoted by $h_i$, and the on-off state of the active excitation at time $t$ for the $i$-th actuator is represented by $a_i(t)$.
Here $i=1,2,...,N$ corresponds to the index of the corresponding multi-modal actuator.

\subsection{Optimization problem formulation}\label{ssec:S2.2}
One of the main objectives in this study is to optimize passive and active flow control laws and minimize flow separation in the Smart Skin experimental platform.
In this study we focus on the open-loop control and the flow control laws can be parameterized as follows.
our focus is on the open-loop control, and the flow control laws can be parameterized as follows.
For passive control, we discretize the variable $h_i$ into five possible heights, allowing each multi-modal actuator to choose from \SI{0}{\milli\meter}, \SI{2}{\milli\meter}, \SI{4}{\milli\meter}, \SI{6}{\milli\meter}, and \SI{8}{\milli\meter}.
For active control, we enable constant blowing, and the active parameter $a_i\in\{0,1\}$ represents the on-off status of the $i$-th actuator.

To cooperate with the optimization algorithms, we define a cost function $J_a$ based on the pressure distribution on the ramp surface.
To reduce separation and acheive the maximum pressure recovery on the ramp surface, we mathematically define the cost function as:
\begin{equation}
J_a =  \int(\bar{p}_{0}-\bar{p})dS.
    \label{eqn:eqn1}
\end{equation}
In this equation, $\bar{p}_0$ represents the pressure in the freestream, and $\bar{p}$ is the time-averaged pressure distribution on the ramp surface $S$.
By formulating the flow control problem in this manner, we transform the original challenge into a parameter optimization problem under a minimization framework.
To better understand the effect of control laws during optimization, we introduce a non-dimensionalized version of the cost function (denoted as $J_a^*$) for the discussions in \Cref{sec:S3}. This non-dimensionalized version of the cost function is based on the ratio between $J_a$ under a certain actuation pattern and the baseline cost function $J_{a,\text{baseline}}$ obtained under the unforced case. 
Mathematically $J_a^*$ is formulated as:
\begin{equation}
J_a^* =  1-\frac{J_a}{J_{a,\text{baseline}}}.
    \label{eqn:eqn2}
\end{equation}
In this non-dimensionalized version, a positive value of $J_a^*$ indicates a negative pressure recovery, implying that the flow field has deteriorated. Conversely, a negative value of $J_a^*$ signifies an improvement in flow separation control.
The parameterization of the control strategies and the definition of the cost function in the Smart Skin plant leads to a 60 dimensional optimization problem.
As the key focus of this study, we aim to address this challenging problem by leveraging a fast-converging machine learning optimization algorithm.

\subsection{PSO-TPME: a fast-converging optimizer through targeted, position-mutated, elitism}\label{ssec:S2.3}

%%%%%%%%%%%%%BEGIN %FIGURE%%%%%%%%%%%%%%%%
\begin{figure}
\centering
\includegraphics[width=.95\linewidth]{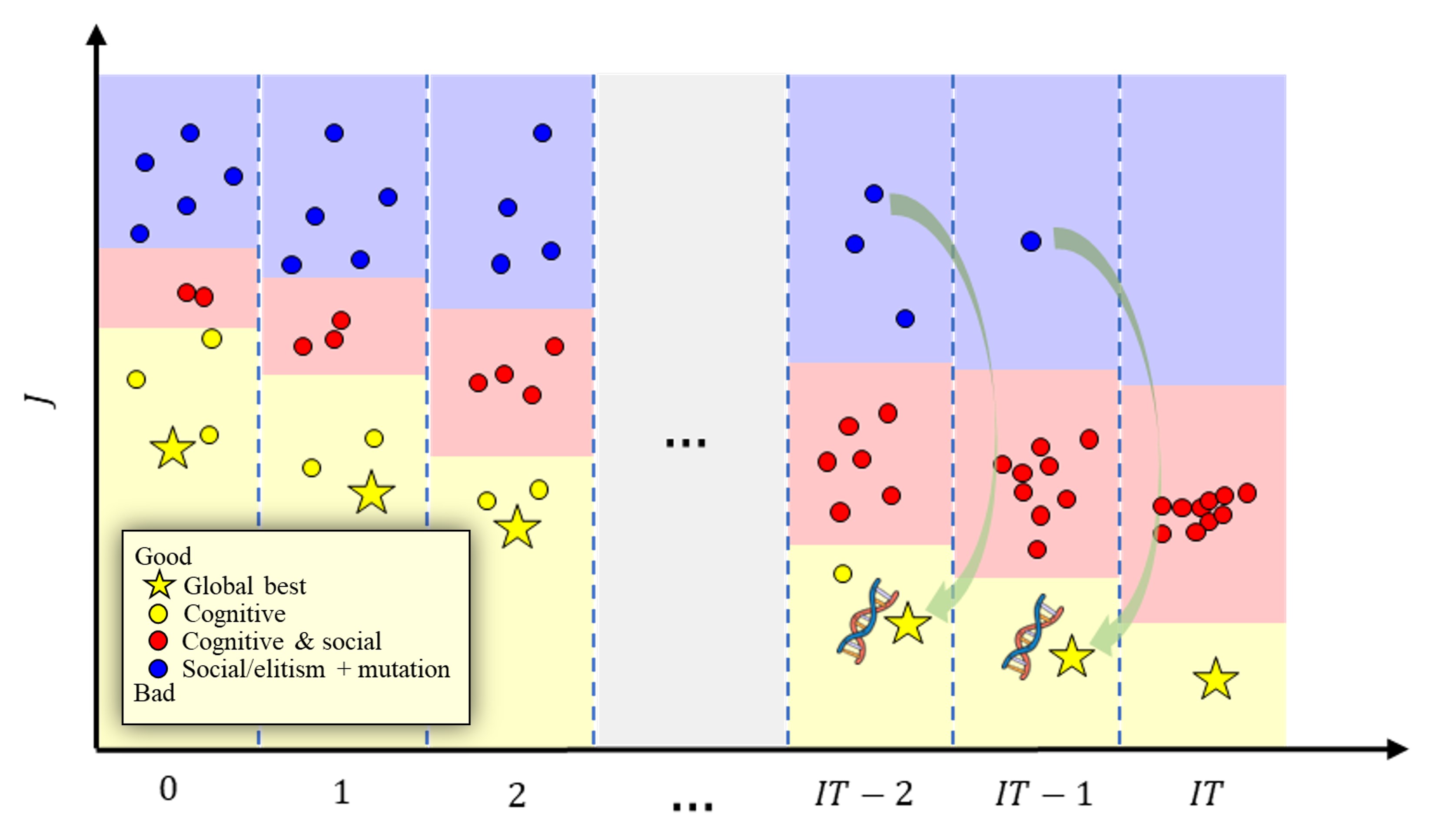}
\caption{A graphic illustration of the particle swarm optimization with targeted, position-mutated, elitism (PSO-TPME, \citealt{Shaqarin2023}).}
\label{fig:pso}
\end{figure}
%%%%%%%%%%%%%%%END %FIGURE%%%%%%%%%%%%%%%%

To solve the above-mentioned optimization problem, we employ a cutting-edge variant of the particle swarm optimization (PSO-TPME, \citealt{Shaqarin2023}).
A visual representation of the PSO-TPME algorithm is depicted in \Cref{fig:pso}. 
PSO is a widely recognized evolutionary optimization technique that mimics the collective behavior observed in social systems, such as bird swarms. 
The PSO-TPME variant builds upon the traditional PSO approach by incorporating targeted position mutation and elitism, leading to significantly improved convergence rates and enhanced accuracy compared to other PSO variants.
In PSO-TPME, the particles, which represent candidate solutions, are dynamically categorized into three groups based on their performance: good, fair, and bad. 
Each category is assigned specific exploration and exploitation characteristics. 
Good particles are empowered with enhanced exploitation capabilities within the local domain. 
For bad particles, a position-mutated elitism process is activated to accelerate the convergence rate. 
Fair particles will evolve according to the standard PSO procedure.
This cutting-edge PSO optimization algorithm can effectively tackle high-dimensional optimization problems, and has demonstrated successful application in power-voltage systems \citep{Shaqarin2023_b}.
For further details and comprehensive information about this algorithm, please refer to \citet{Shaqarin2023}.

\subsection{Experimental setup and instrumentation}\label{ssec:S2.3}
% optimization experiment
The key parameters of the flow control optimization experiment are listed in \Cref{tab:pso}.
We simultaneously optimize both passive and active parameters with PSO-TPME.
It is important to note that while the plant parameters are discrete values, the PSO-TPME algorithm operates in a continuous space. 
Therefore, the parameters within each particle are rounded to the nearest integer to determine the corresponding action.
In addition, the active control for an actuator with a specified height of 0 won't be executed.
The number of particles was set to 35, which accounts for approximately 58 \% of the total number of parameters. 
For each particle candidate, surface pressure data is collected for 10 seconds at a frequency of 1024 Hz after the corresponding actions of the actuators.
The evolutionary process of the PSO algorithm consists of 1000 iterations, resulting in an overall training time of approximately 15 hours.
We perform five independent runs to assess the statistical performance of the optimizer.
Detailed analysis of the optimization results will be presented in \Cref{ssec:S3.3}.

%%%%%%%%%%%%%TABLE%%%%%%%%%%%%%%%%
\begin{table}
	\begin{center}
	\def~{\hphantom{0}}
	\begin{tabular}{P{5cm} | P{5cm} }
        Parameter & Value\\ \hline
        Number of parameters & 30 (passive) + 30 (active)\\
        Passive parameters & $h_i = \{0,1,2,3,4\}\times \SI{2}{\milli\meter}$ \\
        Active parameters & $a_i = \{0,1\}$ \\
        Number of particles ($P$) & 35\\
        Number of Iterations ($IT$)& 1000\\
        Sampling frequency&\SI{1024}{\hertz}\\
        Testing time for each individual&\SI{10}{\second}\\
        Number of independent runs & 5\\

	\end{tabular} 
	\caption{Experimental setup for the Smart Skin control optimization using PSO-TPME.}
	\label{tab:pso}
	\end{center}
\end{table}
%%%%%%%%%%%%%%%TABLE%%%%%%%%%%%%%%%%

% HWA
In addition to the flow control optimization experiment, several experimental techniques are employed to characterize the baseline and controlled flow field. 
One of these techniques involves the use of hotwire measurements to obtain the boundary layer profiles at the leading edge of the ramp profile ($x=0$).
The hotwire system used in this study is a single-wire sensor and a CTA-02 anemometry from HangHua Inc. 
By employing a traverse system, we utilize the hotwire probe to measure the boundary layer profile from  $z=\SI{1}{\milli\meter}$ to $\SI{30}{\milli\meter}$. 
The hotwire anemometry is connected to an NI USB-6009 for data acquisition.
The sampling frequency is set to \SI{2048}{\hertz}, and the measurement time for each location is \SI{20}{\second}.
To ensure accurate velocity readings, the hotwire anemometry is calibrated by recording the output voltage of the hotwire at various incoming velocities. 
A fifth-order polynomial fitting is then applied to establish the relationship between flow velocities and mean output voltages.

% PIV
In addition, particle image velocimetry (PIV) is employed to measure the velocity fields on seven different streamwise planes. 
These measurement planes are aligned with the spanwise location of the pressure taps, providing valuable insights into the flow field under the baseline case and the optimal control laws, respectively.
A planar PIV configuration is adopted to obtain two-dimensional velocity measurements.
Two cameras are positioned side-by-side to maximize the field of view (FoV) in the streamwise direction.
The seeding particles are generated by an Antari fog machine with an average size of approximately \SI{0.2}{\micro\meter}.
A dual-cavity Nd:YAG laser with a wavelength of \SI{532}{\nano\meter} is employed to illuminate the seeding particles. 
For each measurement plane, a total of 3000 image snapshots are acquired at a sampling rate of \SI{12}{\hertz}. 
The acquired image pairs are subsequently post-processed to obtain the velocity field. 
A standard multi-pass routine is adopted to calculate velocity vectors, where the final pass utilizes an interrogation window size of $32\times 32$ with 50\% overlap. 
This configuration ensures a spatial resolution of approximately 0.5 mm for all measurements.
The uncertainty analysis is performed based on the cross-correlation statistics during the calculation of velocity vectors \citep{Wieneke2015}. 
This analysis suggests an uncertainty less than \SI{0.1}{\meter\per\second} for all PIV measurement campaigns.
\section{Results and discussions}\label{sec:S3}

\subsection{Characterization of the unforced flow}\label{ssec:S3.1}

%%%%%%%%%%%%%BEGIN %FIGURE%%%%%%%%%%%%%%%%
\begin{figure}
\centering
\includegraphics[width=.85\linewidth]{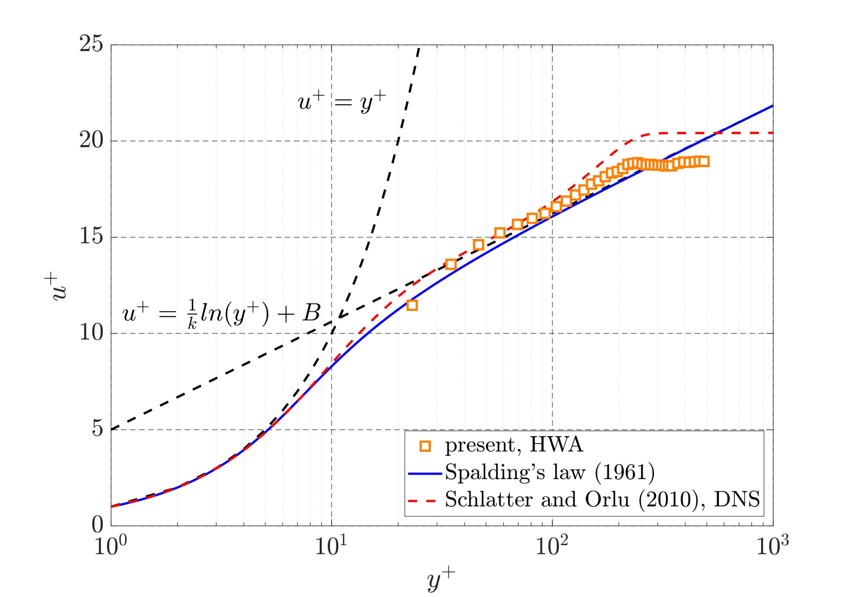}
\caption{Boundary layer profile at the leading edge of the ramp and $z=0$ under the unforced case. The theoretical results from \citet{Spalding1961} and the DNS results from \cite{Schlatter2010} at a similar $Re_{\theta}$ are also presented for a comprehensive comparison.}
\label{fig:BL}
\end{figure}
%%%%%%%%%%%%%%%END %FIGURE%%%%%%%%%%%%%%%%

We start with the examination of the initial flow boundary layer, which is closely related to the ramp flow separation downstream according to \citet{Hildebrand2020}
The boundary layer profile at the ramp leading edge ($x=0$) is measured using the hotwire anemometry and a traverse system to enable scanning in the wall-normal direction.
\Cref{fig:BL} shows the streamwise velocity profile in the inner scaling (i.e., the ``law of the wall'' region).
For a comprehensive comparison, the theoretical solution of \citet{Spalding1961} with standard parameters $B = 5.0$ and $k = 0.41$ is also visualized in this figure. 
The non-dimensionalized velocity profiles confirm the turbulent state of the boundary layer, with a logarithmic layer extending out to about $y^+=230$.
This boundary layer profile suggests a momentum boundary layer thickness of $\theta=\SI{0.98}{\milli\meter}$,
corresponding to a Reynolds number of $Re_\theta\approx 500$ based on the momentum thickness.
A very good agreement was found between the experimental profile and the boundary layer profile at $Re_\theta = 670$ taken from the DNS simulations of in \citet{Schlatter2010} a zero-gradient boundary layer. 
As the minimum distance between the hotwire sensor and the wall is set to $\SI{1}{\milli\meter}$, the region below $y^+=23$ can not be resolved from the hotwire measurements.
Therefore, the viscous sublayer, which is typically below $y^+=10$, can not be effectively resolved from the current set of hotwire measurements.

%%%%%%%%%%%%%BEGIN %FIGURE%%%%%%%%%%%%%%%%
\begin{figure}
\centering
\includegraphics[width=.9\linewidth]{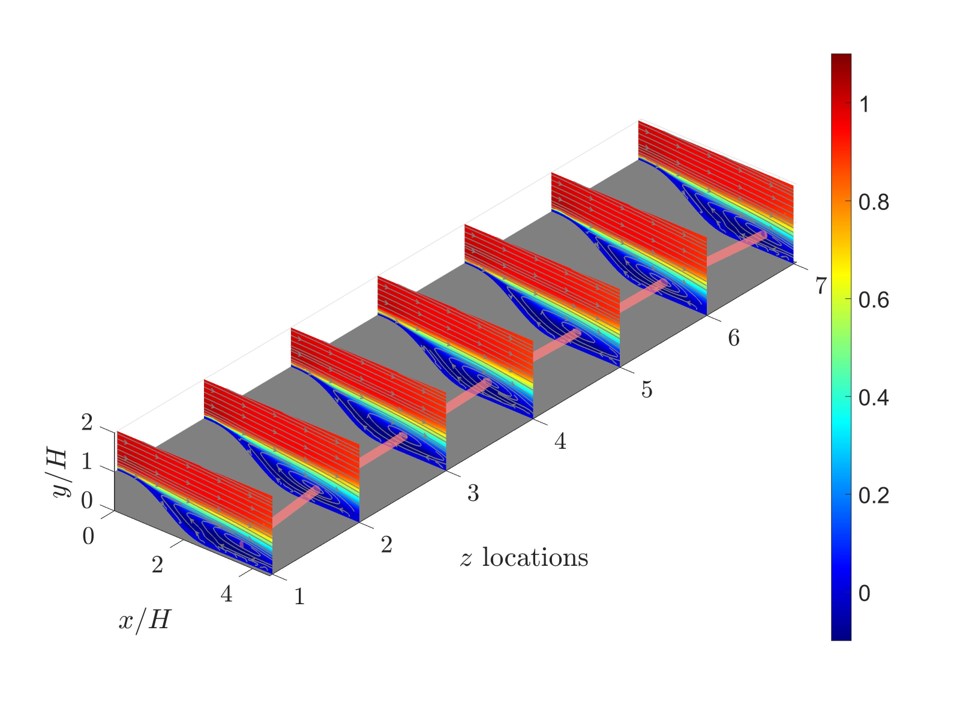}
\caption{Streamwise velocity contours and the superpositioned 2D streamlines measured at 7 spanwise locations with PIV. The spanwise locations are aligned with the locations of the pressure taps. The $z$ locations 1 to 7 correspond to $z/D=$ - 1.8, 1.2, 0.6, 0, 0.6, 1.2, and 1.8, respectively. The red tube connects the vortex core locations on each measurement plane according to the method described in \citet{Graftieaux2001}.}
\label{fig:3Dbaseline}
\end{figure}
%%%%%%%%%%%%%%%END %FIGURE%%%%%%%%%%%%%%%%

To investigate the characteristics of flow separation over the Smart Skin in unforced conditions, a series of 2D PIV measurements are performed in this study.
\Cref{fig:3Dbaseline} presents a comprehensive three-dimensional visualization of the velocity fields obtained at 7 different spanwise locations. 
These measurement planes align with the locations of the pressure taps in the Smart Skin configuration (refer to \Cref{fig:SS}), and the spacing between consecutive streamwise measurement planes is at $\Delta z = 0.6H$. 
In addition, 2D streamlines are overlaid on each measurement plane to enhance the understanding of flow separation and recirculation phenomena.
Strong similarities can be observed among the velocity profiles, including the expansion of separated shear layers, as well as the occurrence of flow recirculations below.
To obtain the vortex core in the recirculation region, a vortex identification method proposed in \citet{Graftieaux2001} is applied to the velocity field on each measurement plane, and the identified vortex locations are connected by a red thick line in the figure.
The distribution of vortex cores demonstrates strong symmetry about $z=0$, with the locations of vortex cores shifting further downstream when the measurement planes approach the side walls. 

%%%%%%%%%%%%%BEGIN %FIGURE%%%%%%%%%%%%%%%%
\begin{figure}
\centering
\includegraphics[width=.6\linewidth]{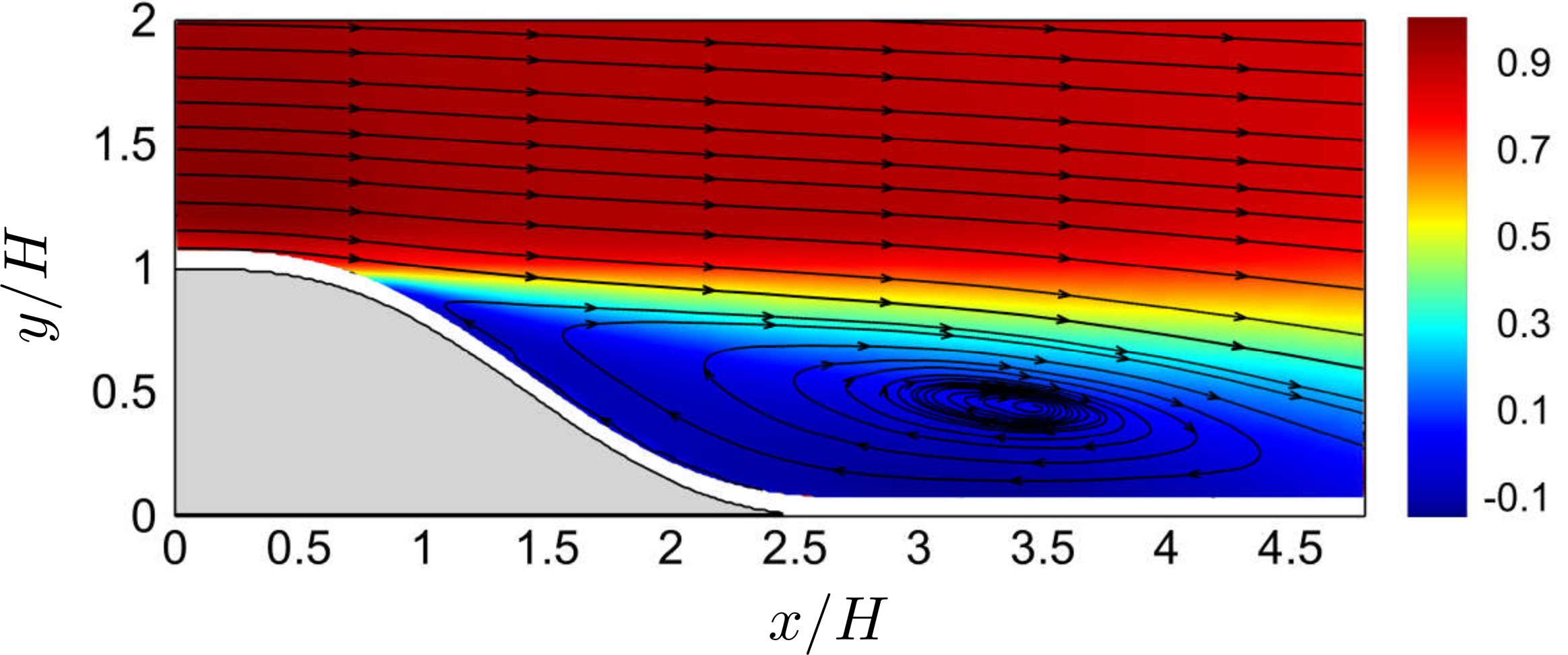}
\caption{Mean streamwise velocity $U_x$ superpositioned with 2D streamlines at $z=0$ obtained from PIV measurements.}
\label{fig:mean_CL}
\end{figure}
%%%%%%%%%%%%%%%END %FIGURE%%%%%%%%%%%%%%%%

%%%%%%%%%%%%%BEGIN %FIGURE%%%%%%%%%%%%%%%%
\begin{figure}
\centering
\includegraphics[width=.6\linewidth]{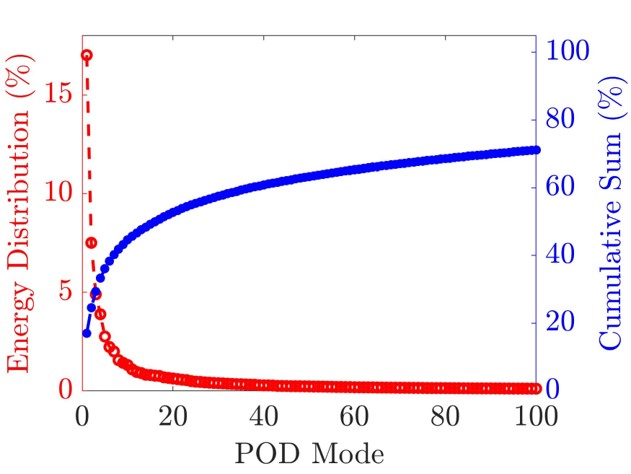}
\caption{Energy distribution of the first 100 POD modes (red) and the cumulative sum (blue).}
\label{fig:POD_energy}
\end{figure}
%%%%%%%%%%%%%%%END %FIGURE%%%%%%%%%%%%%%%%

In addition, we examine the turbulence characteristics of the flow using the proper orthogonal decomposition (POD, \citealt{Lumley1967}) using the recorded velocity snapshots.
As strong similarities are exhibited among different spanwise directions, in the following we only present the results from velocity snapshots measured on the center line of the model ($z=0$).
In this work, we follow the snapshot version of the POD \citep{Sirovich1987}, where the velocity snapshot $\bm{u}(x,y,t_n)$ measured at time instance $t_n$ can be expanded as the ensemble-averaged mean velocity $\bm{U}$, and the linear combinations of orthonormal basis functions $(\bm{\phi}^{(m)}(x,y))$ and the corresponding modal coefficients $a^{(m)}(t_n)$:
\begin{equation}
\bm{u}(x,y,t_n) =  \bm{U}(x,y)  + \sum_{m=1}^{M} \bm{\phi}^{(m)}(x,y)a^{(m)}(t_n).
\end{equation}
In this equation, $m$ represents the $m$-th POD mode and $M$ represents the truncation of the leading POD modes.
A close-up view of the mean streamwise velocity field $\bm{U}_x$ and the two-dimensional streamlines is presented in \Cref{fig:mean_CL} for reference.
The contribution of the first 100 POD modes to the turbulent kinetic energy, as well as the cumulative sum, are presented in \Cref{fig:POD_energy}.
The first POD mode captures approximately 17 \% of the turbulent kinetic energy (TKE), while the second and third POD modes only contribute to 7.5 \% and 5 \% of the TKE, respectively.
Similar observations are also reported in \citet{Kourta2012} for a straight ramp profile and the rear slant of an Ahmed body in \citet{Thacker2013}, in which the first POD mode in the separated flow is believed to be associated with the flapping motion of the recirculation region in the wall-normal direction while the higher order modes relate to the convective flow structures in the shear layer.

%%%%%%%%%%%%%BEGIN %FIGURE%%%%%%%%%%%%%%%%
\begin{figure}
\centering
\includegraphics[width=.99\linewidth]{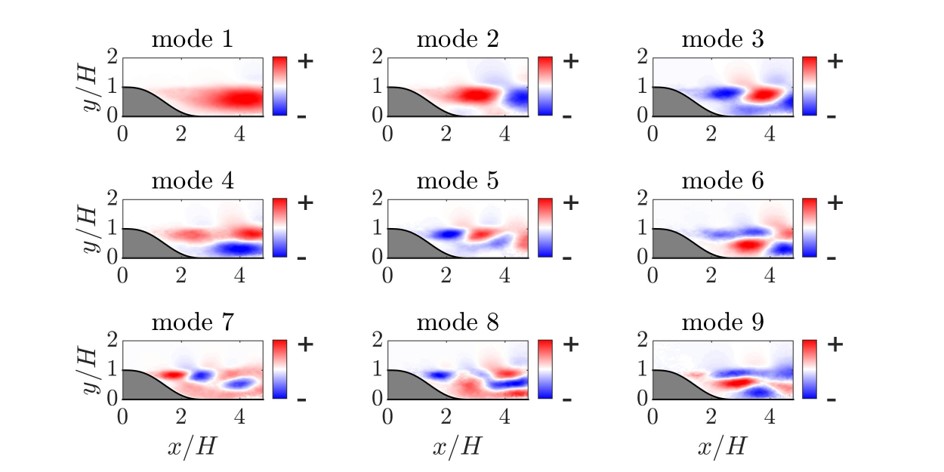}
\caption{The streamwise component of the leading POD modes.}
\label{fig:POD_u}
\end{figure}
%%%%%%%%%%%%%%%END %FIGURE%%%%%%%%%%%%%%%%

%%%%%%%%%%%%%BEGIN %FIGURE%%%%%%%%%%%%%%%%
\begin{figure}
\centering
\includegraphics[width=.99\linewidth]{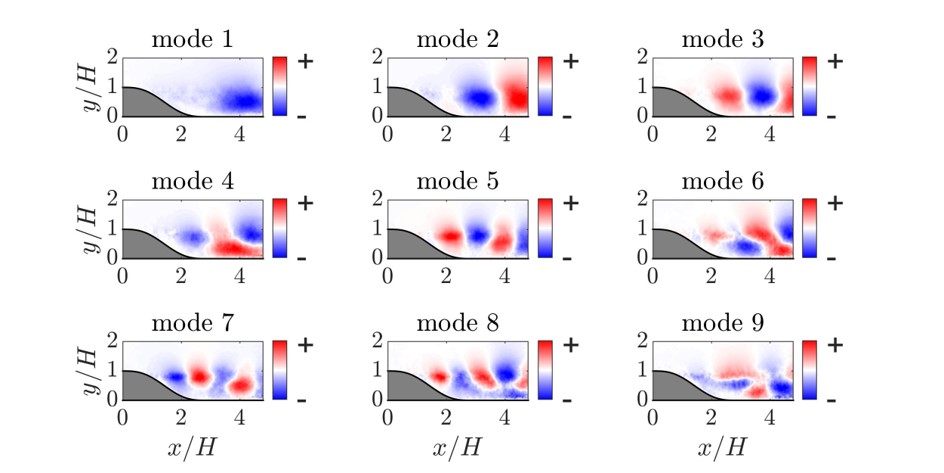}
\caption{The wall-normal component of the leading POD modes.}
\label{fig:POD_v}
\end{figure}
%%%%%%%%%%%%%%%END %FIGURE%%%%%%%%%%%%%%%%

\Cref{fig:POD_u,fig:POD_v} presents the streamwise and wall-normal components of the first 9 POD modes.
The first POD mode corresponds to large-scale fluctuations inside the recirculation region. 
A thorough discussion of the first POD mode in the ramp flow is presented in \citet{Kourta2015}, in which the low-order reconstruction combining the mean velocity and the first POD mode demonstrates the flapping motion of the recirculation region.
Modes 2 and 3 correspond to a pair of wave-like turbulent structures in the flow, and the combination of these two modes can represent turbulence convection at different phases.
The turbulent structures in these two modes  are of the largest scale, and the turbulent structures primarily reside inside the shear layer.
The higher-order modes also feature alternating positive/negative structures inside the shear layer, although the turbulent strucutres becomes smaller in size. 
The above-mentioned turbulent mechanisms, including the vertical flapping motion of the recirculation region, as well as the downstream convection of large-scale coherent structures inside the separated shear layer, are highly consistent with the discussions in \citet{Cherry1984,Kiya1985,Largeau2006}, among many others.

%%%%%%%%%%%%%BEGIN %FIGURE%%%%%%%%%%%%%%%%
\begin{figure}
\centering
\includegraphics[width=.7\linewidth]{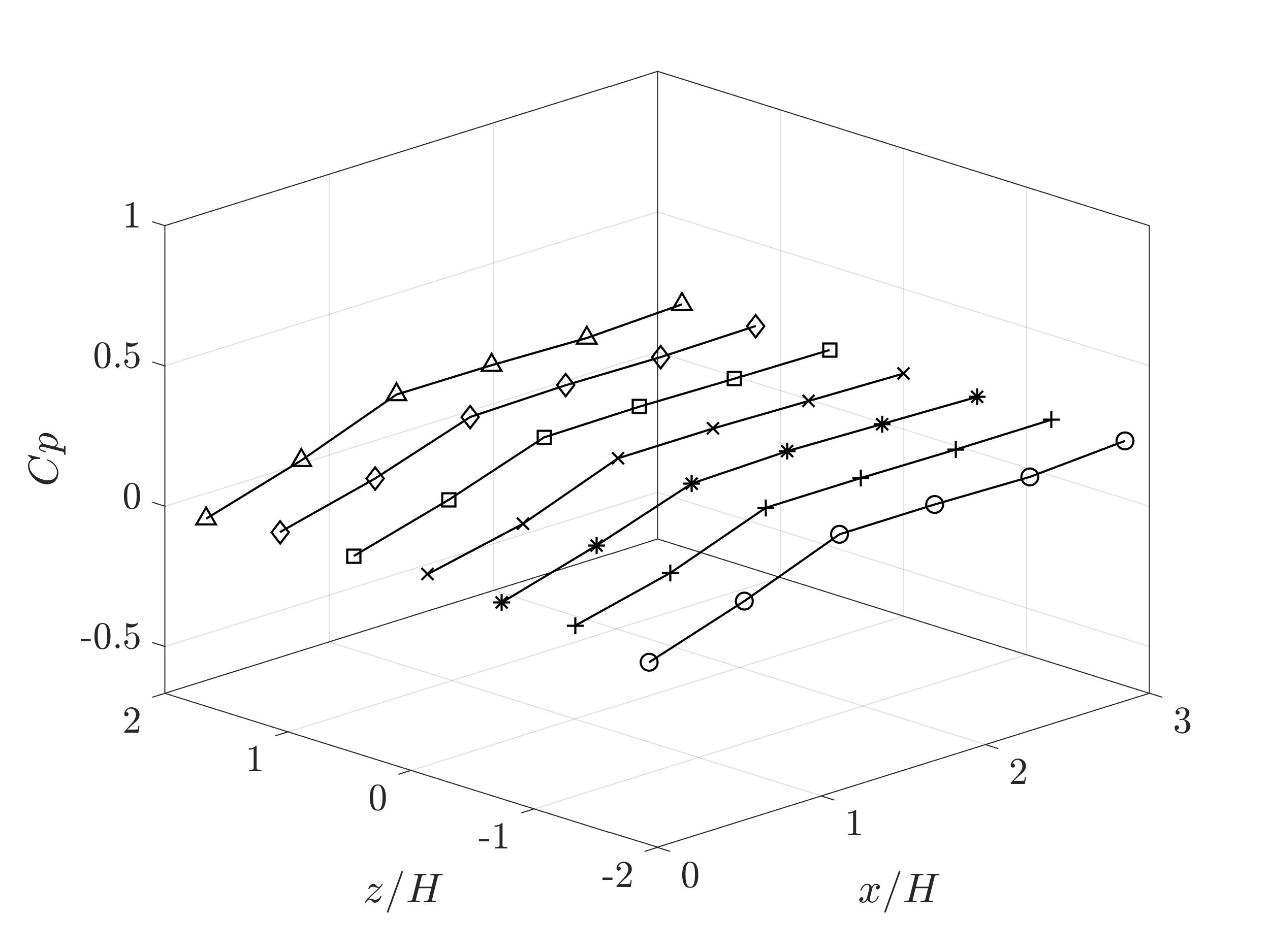}
\caption{Pressure distribution on the ramp surface under the unforced case.}
\label{fig:Cp_baseline}
\end{figure}
%%%%%%%%%%%%%%%END %FIGURE%%%%%%%%%%%%%%%%
The distribution of the pressure coefficient ($Cp$) on the ramp surface is presented in \Cref{fig:Cp_baseline}.
The calculation of the pressure coefficient is based on the pressure measurements at the pressure tap locations.
Strong similarities can be observed for the pressure profiles at all spanwise locations.
The pressure coefficients near the ramp leading edge are close to 0 at all spanwise locations.
Downstream of the ramp, the pressure rises up to the maximal value ($Cp \approx 0.2$) from $x/H=1.26$ to $x/H=3$.
By reducing the pressure separation on the ramp, the downstream surface pressure is expected to be further enhanced with distributed multi-modal actuators.

\subsection{Parametric study on multi-modal flow control using the Smart Skin}\label{ssec:S3.2}
This subsection presents the parametric study of the flow control outcomes using the Smart Skin with distributed, multi-modal actuators.
We choose a total of 120 actuation control parameters and examine the relative performance regarding the pressure recovery on the ramp surface.
These actuation patterns include purely passive and passive + active control commands.
In addition, the pressure distribution, as well as the flow fields from PIV measurements are presented, and compared with the unforced case.
By comparing the flow control performance, we would like to stress the difficulty of selecting the proper flow control strategy with purely human experience, which facilitate the need for machine learning optimization of the control strategy.

%%%%%%%%%%%%%BEGIN %FIGURE%%%%%%%%%%%%%%%%
\begin{figure}
\centering
\includegraphics[width=.99\linewidth,trim={2cm 0 2cm 0},clip]{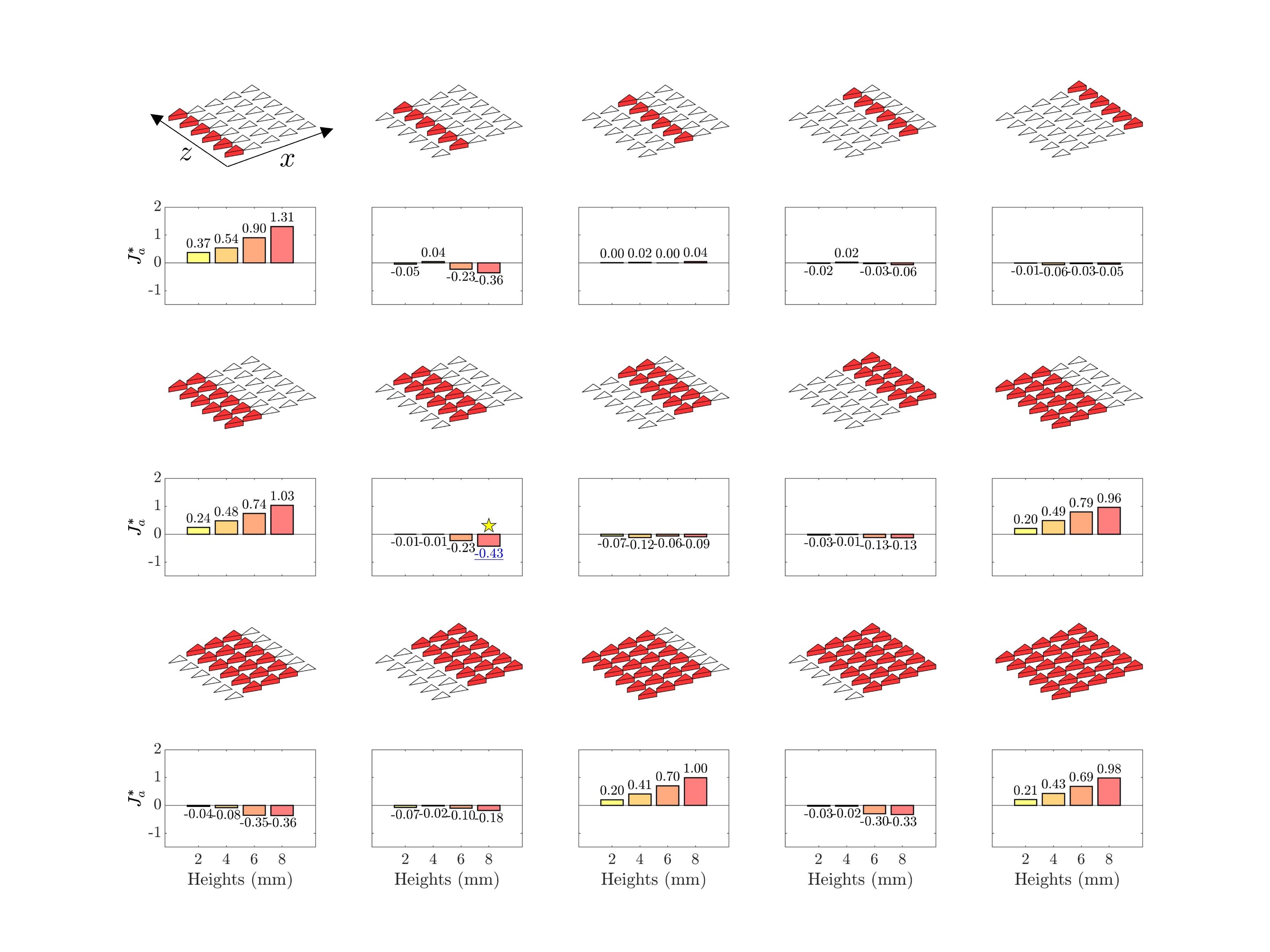}
\caption{The parametric results of different passive control laws. The activated multi-modal actuators are  operated under a passive-only mode and are denoted in red.
A total of 4 different heights are evaluated for each combination of activated actuators. The best flow control result is denoted by a yellow star in the figure.}
\label{fig:parametric1}
\end{figure}
%%%%%%%%%%%%%%%END %FIGURE%%%%%%%%%%%%%%%%

\Cref{fig:parametric1} presents the parametric results when the actuators are operated under the passive-only cases.
The parametric patterns are selected such that each spanwise row operates in a uniform manner.
We tested the performance of single-row, double-row, ..., and all-row actuations.
For each case, the selected actuators, which are denoted in red in the figure, are lifted up at four different heights (\SI{2}{\milli\meter}, \SI{4}{\milli\meter}, \SI{6}{\milli\meter}, \SI{8}{\milli\meter}) to examine the corresponding flow control performance ($J_a^*$).
As introduced in \Cref{ssec:S2.2}, a negative $J_a^*$ represents the improved pressure recovery on the ramp surface, while a positive $J_a^*$ represents the increased pressure loss.
Among the single-row actuation results, the activation of the second row at the $\SI{8}{\milli\meter}$ is the most effective flow control strategy, and can lead to a 36 \% pressure recovery compared to the baseline case.
Conversely, activating the first row at higher actuator heights results in increased pressure loss. 
In other cases, the single-row actuation patterns will lead to negligible flow control improvements.
Interestingly, similar observations can be found in multi-row actuation results.
The flow control patterns starting with the activation of the second row will always lead to improved flow control performances as the actuation height increases.
However, the flow patterns involving the first row will result in increasingly worse flow control outcomes as the height increases.
Without the inclusion of the first and second rows, the individual or combined actuation of the downstream rows can hardly influence the flow control performance under all cases.
The best flow control performance is observed when the second and the third rows are lifted up to the maximum height of \SI{8}{\milli\meter}, where the generation of counter-rotating vortices is believed to improve the flow separation and recovers the ramp surface pressure by 43 \% compared to the baseline case.

%%%%%%%%%%%%%BEGIN %FIGURE%%%%%%%%%%%%%%%%
\begin{figure}
\centering
\includegraphics[width=.99\linewidth,trim={2cm 0 2cm 0},clip]{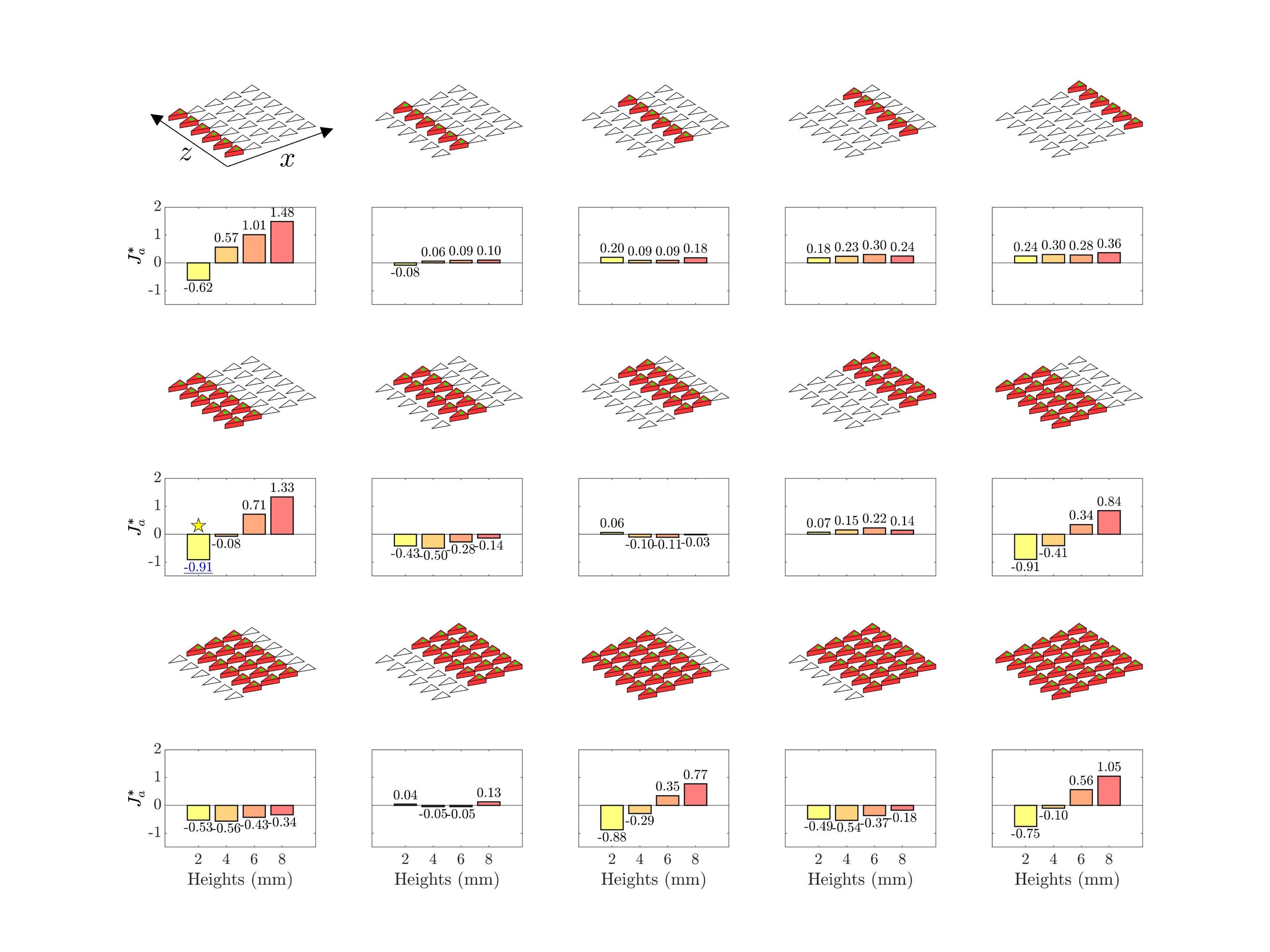}
\caption{The parametric results of different passive + active control laws. 
The activated multi-modal actuators are  operated under the combined passive and active mode and are denoted in red, and the green dots on top of these actuators represent the active control.
A total of 4 different heights are evaluated for each combination of activated actuators. The best flow control result is denoted by a yellow star in the figure.}
\label{fig:parametric2}
\end{figure}
%%%%%%%%%%%%%%%END %FIGURE%%%%%%%%%%%%%%%%

The impact of multi-modal actuation can be further assessed by incorporating additional active control alongside the passive-only patterns. 
\Cref{fig:parametric2} displays the flow control results when the actuators execute both passive and active controls. 
We keep the activation patterns investigated in \Cref{fig:parametric1}, and allow additional active actuation in each case.
Here, the active control is the constant blowing of the minijet actuators in the streamwise direction, and the involvement of time-periodic control laws will be investigated in future work.
For clarity, the activation of active control is denoted by green dots in the figure.
Concerning single-row actuation patterns, notable pressure recovery is observed when the first row is blowing at a minimum height of \SI{2}{\milli\meter}. 
However, as the actuator height increases, the performance of first-row blowing progressively worsens. 
While the activation of the second row leads to significant pressure recovery in the passive-only scenario, negligible flow control effects are observed at all heights when additional active control is introduced. 
The individual control patterns for the remaining single-row actuations result in additional pressure losses compared to the baseline case.
In the case of flow control involving multiple rows, including the first row at \SI{2}{\milli\meter} can  consistently produce promising results. 
Furthermore, the best flow control outcome is observed when the first two rows are activated at 2 mm, yielding a remarkable 91 \% pressure recovery compared to the baseline case. 
These findings suggest that active control from the first row at the minimum height consistently enhances the flow field. 
However, as the actuator height increases, the effectiveness of flow control diminishes. 
Multi-row control patterns that begin with the second row also ensure a guaranteed pressure recovery at various heights. 
Nevertheless, actuation patterns excluding the first two rows only yield negligible effects on ramp wall pressure recovery.

By comparing parametric evaluations in \Cref{fig:parametric1} and \Cref{fig:parametric2}, we can obtain an enhanced understanding to the effect of passive and active actuations in the Smart Skin. 
The inclusion of active control introduces notable changes in the performance of previously effective actuation patterns in the passive-only case, and vice versa. 
For example, the effects of the first and second rows differ significantly between the passive-only and passive+active modes.
These observations imply that it is difficult to isolate the impact of active and passive control when designing an effective flow control strategy. 
In addition, as we include more parametric patterns in flow control evaluations, more complex dynamics may emerge. 

%%%%%%%%%%%%%BEGIN %FIGURE%%%%%%%%%%%%%%%%
\begin{figure}
\centering
\includegraphics[width=.7\linewidth]{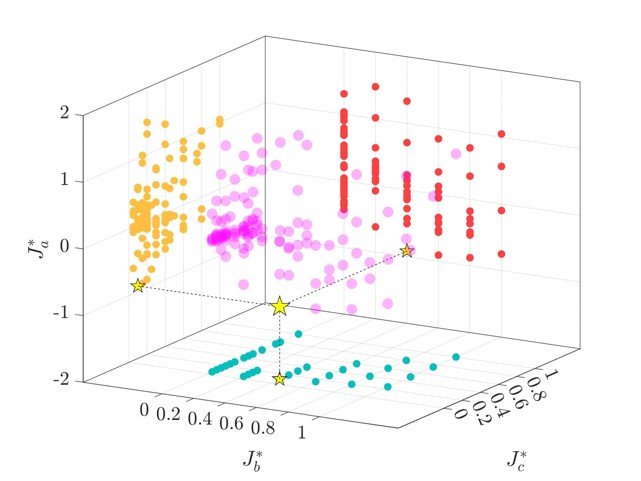}
\caption{Three-dimensional visualization of the relationship between the cost function $J_a^*$, equivalent duty cycle $J_b^*$ and the equivalent actuator height $J_c^*$ from the parametric study. The location of the best performed case is denoted by yellow star.}
\label{fig:3d_pareto}
\end{figure}
%%%%%%%%%%%%%%%END %FIGURE%%%%%%%%%%%%%%%%

In the parametric study, we also define the equivalent duty cycle $J_b^*$ and the equivalent actuator height $J_c^*$.
The definition of the two parameters follows the following two equations:

\begin{equation}
    J_b^* = \langle h_i \rangle_i / h_{\text{max}},
\end{equation}
\begin{equation}
    J_c^* = \langle a_i \rangle_i. 
\end{equation}
In these equations, $\langle \cdot \rangle_i$ represents the ensemble average over all actuators and $h_{\text{max}=\SI{8}{\milli\meter}}$ is the maximum height of the actuators.
\Cref{fig:3d_pareto} displays the relationship between $J_a^*$, $J_b^*$ and $J_c^*$ obtained from the parametric study.
In the visualization of 120 parametric cases, we can observe clear Pareto frontiers from the relationship between $(J_a^*,J_b^*)$ and   $(J_a^*,J_c^*)$ pairs.
For each value of $J_b^*$, the corresponding $J_a^*$ could range from negative values to positive ones, indicating the locations of the active actuators plays a more important role than the number of the actuators.
On the contrary, increasing the equivalent number of actuators can significantly reduce the flow control efficiency. 
The discussions above highlight the need for flow control optimization based on machine learning techniques, and we will show in \Cref{ssec:S3.3} that the utilization of machine learning algorithm can effectively go beyond the frontier and achieve more effective optimization efficiency. 

%%%%%%%%%%%%%BEGIN %FIGURE%%%%%%%%%%%%%%%%
\begin{figure}
\centering
\includegraphics[width=.7\linewidth]{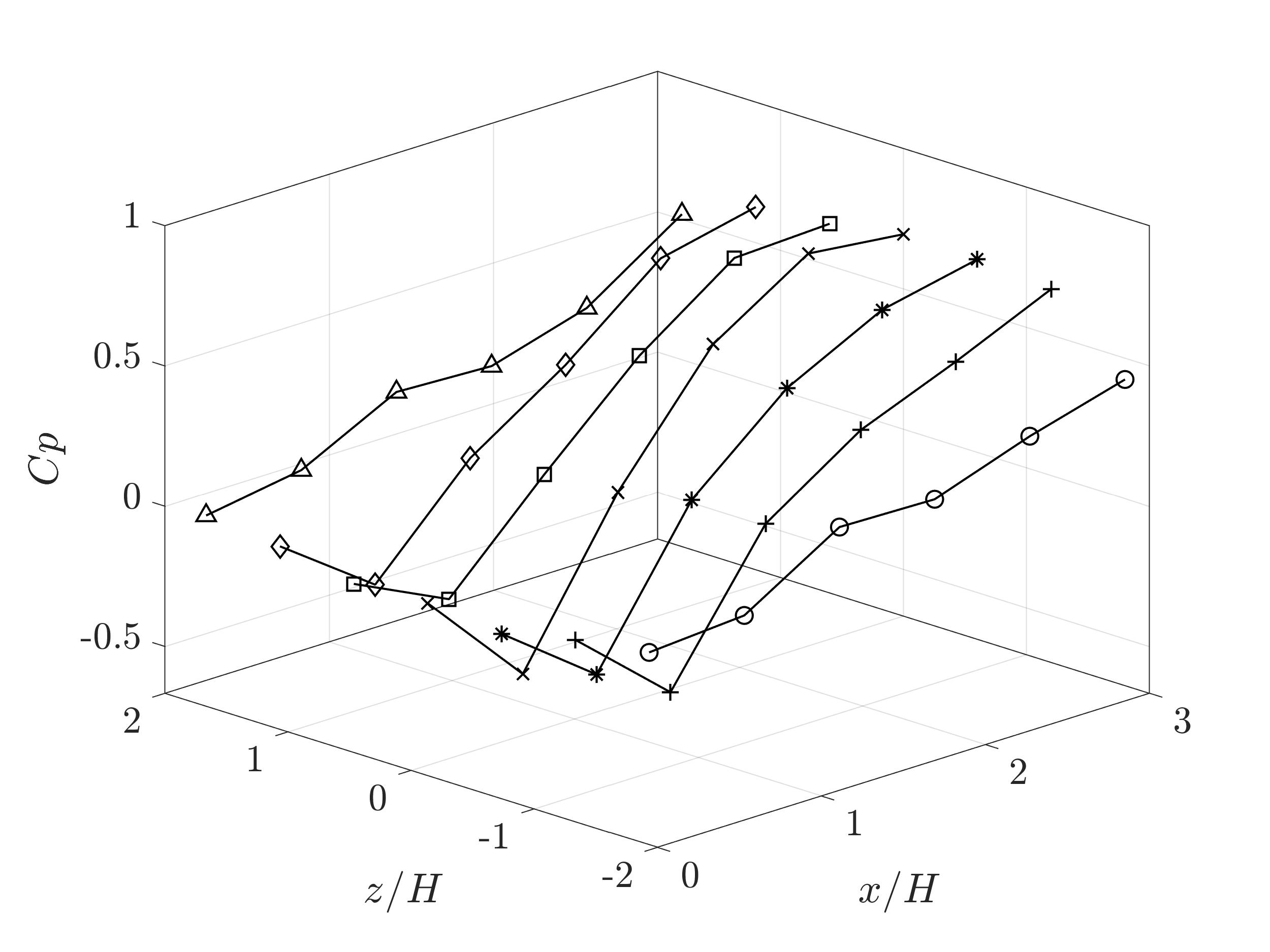}
\caption{Pressure distribution on the ramp surface under the best parametric case.}
\label{fig:pressure_parametric}
\end{figure}
%%%%%%%%%%%%%%%END %FIGURE%%%%%%%%%%%%%%%%

Given the significant flow control outcome observed in the best parametric case (where the first two rows are blowing at a height of \SI{2}{\milli\meter}, see \Cref{fig:parametric2}), flow measurements are conducted based on this specific configuration to further enhance our understanding of the flow control result.
\Cref{fig:pressure_parametric} depicts the profiles of pressure coefficient $C_p$ at a series of spanwise locations. 
Near the leading edge of the ramp, $Cp$ gradually decreases towards the ramp due to a favorable pressure gradient induced by the active blowing.
After the sudden pressure drop, the pressure on the ramp surface starts to continuously recover along the streamwise direction.
At the location $x/H=3$, the pressure coefficient reaches approximately $Cp=0.5$ on the two sides of the model ($z/H=\pm 1.8$), while it reaches about $Cp=0.65$ in the remaining spanwise locations. The relatively lower pressure recovery observed on the two sides is attributed to the absence of corresponding actuators near the side wall.

%%%%%%%%%%%%%BEGIN %FIGURE%%%%%%%%%%%%%%%%
\begin{figure}
\centering
\includegraphics[width=.7\linewidth]{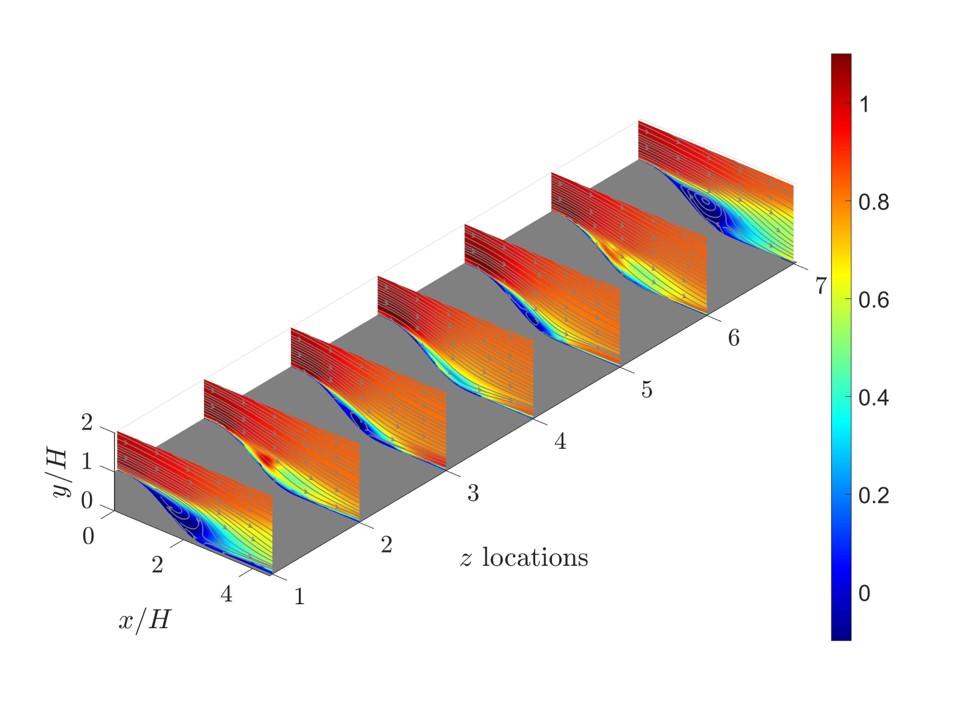}
\caption{Streamwise velocity contours and the superpositioned 2D streamlines measured at 7 spanwise locations with PIV under the best parametric case. The spanwise locations are aligned with the locations of the pressure taps. The $z$ locations 1 to 7 correspond to $z/D=$ - 1.8, 1.2, 0.6, 0, 0.6, 1.2, and 1.8, respectively.}
\label{fig:3Dparametric}
\end{figure}
%%%%%%%%%%%%%%%END %FIGURE%%%%%%%%%%%%%%%%

\begin{figure}
\centering
\includegraphics[width=.8\linewidth,trim={0cm 1.5cm 0cm 1.5cm},clip]{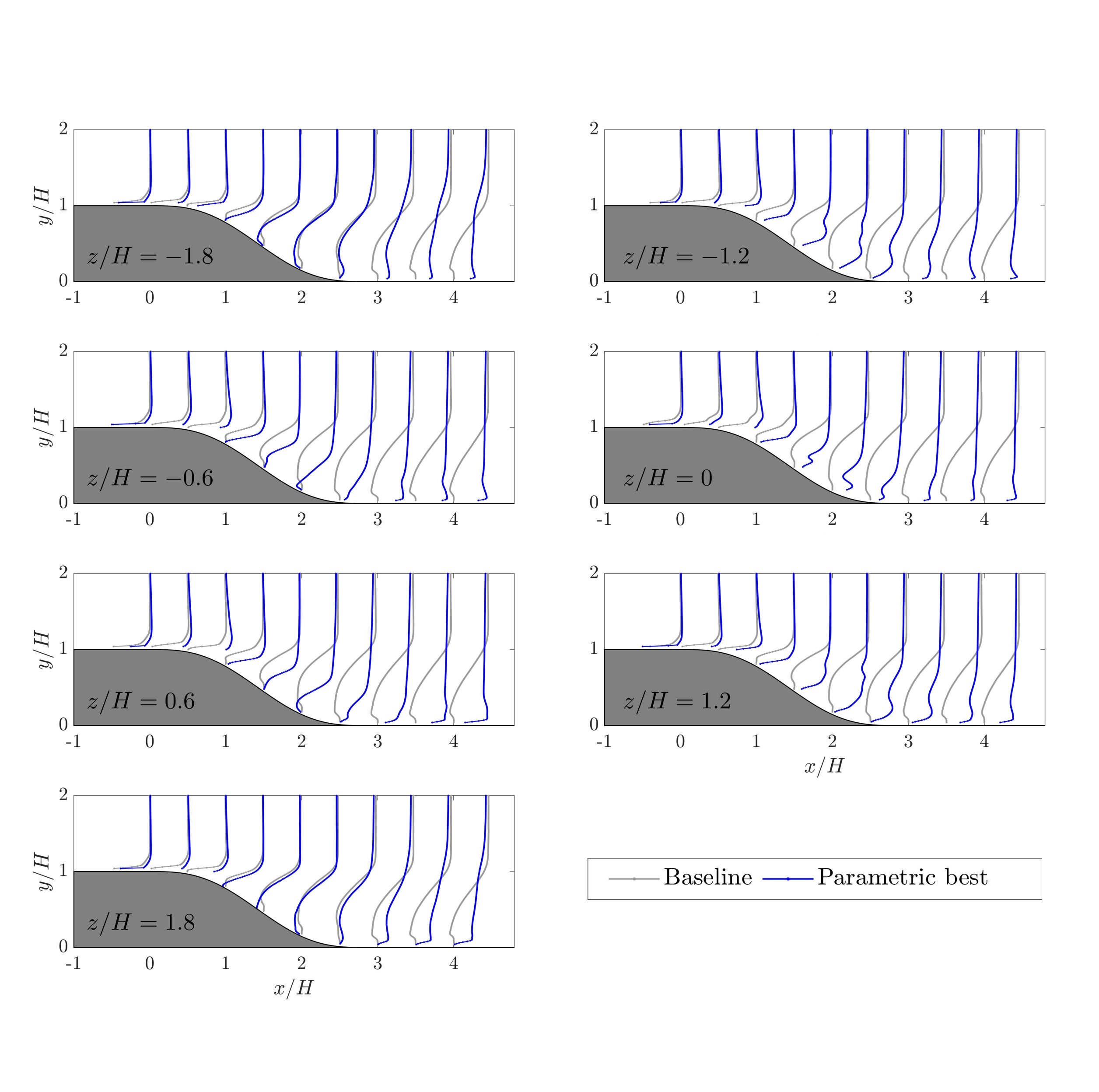}
\caption{Comparison of streamwise velocity profiles at different spanwise locations under the baseline and the best parametric cases.}
\label{fig:2Dparametric}
\end{figure}

\Cref{fig:3Dparametric} presents the streamwise velocity contours and the corresponding 2D streamlines which are measured under the best parametric case.
In comparison to the baseline results shown in \Cref{fig:3Dbaseline}, a significant reduction in flow separation is observed under the best parametric actuations.
At locations $z$ = 2, 4, and 6, flow separation is nearly eliminated, as evidenced by the absence of recirculated streamlines. The low-speed regions concentrate as a very thin layer attached on the ramp wall.
Although small recirculation regions are still observed at locations $z$ = 3 and 5, these regions are thin and short, demonstrating the success of the flow control strategy.
At the near-wall locations 1 and 7, apparent separation regions can still be observed. 
This is primarily due to the absence of corresponding actuators near the wall. Nevertheless, it is evident that the streamwise extent of the recirculation region is reduced with the implementation of single-side actuation.
We also refer to \Cref{fig:2Dparametric} for a more detailed comparison of streamwise velocity profiles under the baseline and the best parametric cases.

%%%%%%%%%%%%%BEGIN %FIGURE%%%%%%%%%%%%%%%%
%%%%%%%%%%%%%%%END %FIGURE%%%%%%%%%%%%%%%%

\subsection{Optimization of the Smart Skin control parameters with PSO-TPME}\label{ssec:S3.3}

%%%%%%%%%%%%%BEGIN %FIGURE%%%%%%%%%%%%%%%%
\begin{figure}
\centering
\includegraphics[width=.65\linewidth]{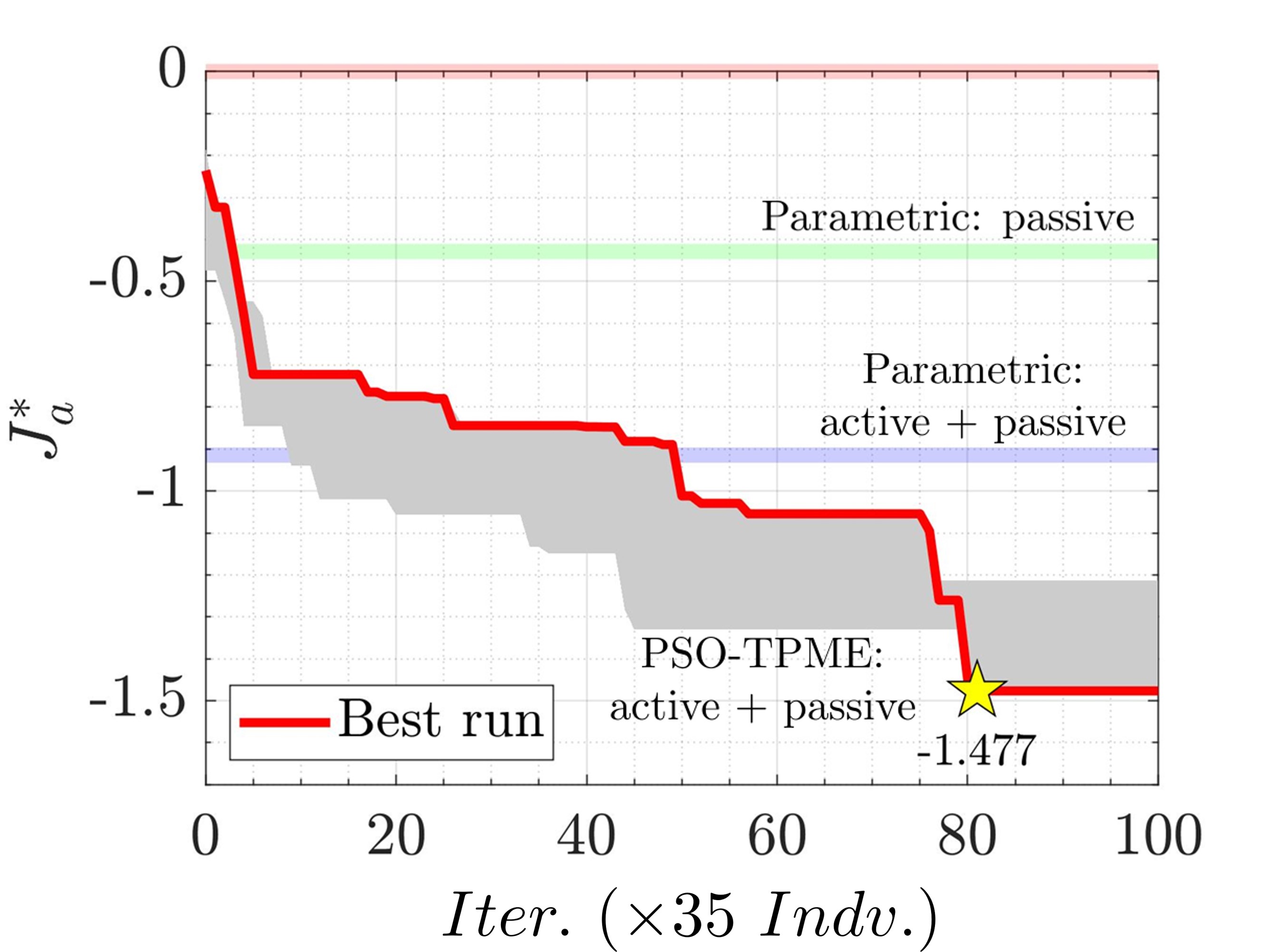}
\caption{Learning curve for the Smart Skin flow control optimization using PSO-TPME.
The shaded area represents the envelope of the lean ring curves from the five individual runs performed in this study.
The red line represents the learning curve of the best run.
The occurrence of the best particle is denoted by a yellow star.
The cost function values of the best passive parametric case and the best active + passive parametric case are also denoted in the figure.}
\label{fig:pso_LC}
\end{figure}
%%%%%%%%%%%%%%%END %FIGURE%%%%%%%%%%%%%%%%

This subsection discusses the machine-learning optimization of the Smart Skin using PSO-TPME.
As introduced in \Cref{ssec:S2.3}, we simultaneous optimize the heights of the multi-modal actuator, as well as the on-off state of the active component.
This challenging 60-parameter optimization experiment is performed within a finite time budget of 15 hours, and a total of 5 independent runs are performed to gain a statistical understanding of the optimizer performance.

\Cref{fig:pso_LC} depicts the learning curve for the Smart Skin flow control optimization achieved using PSO-TPME. 
The learning curve showcases the evolutionary process between the number of iterations and the performance of the currently best particle.
In this work, five independent optimization experiments are conducted. 
The shaded area in the figure represents the envelope of learning curves obtained from these independent runs, while the red line represents the learning curve of the best run.
The performances of the best passive and the best active + passive cases from the parametric study in \Cref{ssec:S3.2} are also displayed in the figure for a comprehensive comparison.
The learning curve envelope in the figure demonstrates the rapid optimization capability of PSO-TPME. 
After only five iterations, the machine-learning optimizer already generates solutions that surpass the performance of the best passive case identified in the parametric study. 
Moreover, the PSO-TPME optimization outperforms the best active + passive case after a total of 50 iterations. 
After 100 iterations, the optimized cost function $J_a^*$ obtained ranges between -1.477 and -1.213.
Although the final outcome in each individual run is slightly different,
it is worth noting that each of these values significantly outperforms the results obtained from the parametric study.  
The slight variation in the optimization outcome can be attributed to the finite time budget imposed by the experimental configuration, 
and the optimization process can only be operated within a constrained time frame. 
The optimization results can be further improved and converged if the overall optimization time is further extended. 
By allowing more time for the optimization process, it is possible to better explore and exploit the search space and potentially achieve better flow control performance.

%%%%%%%%%%%%%BEGIN %FIGURE%%%%%%%%%%%%%%%%
\begin{figure}
\centering
\includegraphics[width=.65\linewidth]{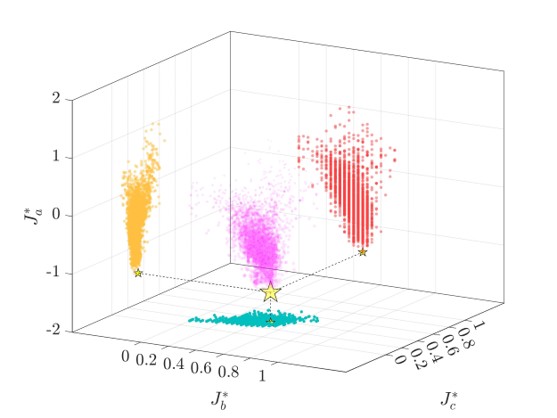}
\caption{Three-dimensional visualization of the relationship between the cost function $J_a^*$, equivalent duty cycle $J_b^*$ and the equivalent actuator height $J_c^*$ from the best run of PSO-TPME optimization. The location of the best parametric case is denoted by yellow star.}
\label{fig:pso_pareto}
\end{figure}
%%%%%%%%%%%%%%%END %FIGURE%%%%%%%%%%%%%%%%

In the subsequent discussion, we will primarily focus on the evolutionary process of the best run, and the performance of the best individual obtained from the corresponding optimization process.
\Cref{fig:pso_pareto} presents the three-dimensional visualization of the relationship between the cost $J_a^*$, as well as the equivalent duty cycle $J_b^*$ and the equivalent actuator height $J_c^*$ for all particles in the best run case.
We choose the same color code as in \Cref{fig:3d_pareto} for a clear comparison between the parametric study and the machine learning optimization.
In contrast to the parametric study where nearly 50 \% of the individuals had $J_a^*$ greater than zero ( i.e., the flow control leads to additional pressure loss), the PSO-TPME optimization demonstrates a more effective performance. 
In the figure, only 6 \% of the particles in the best run case lead to a positive $J_a^*$ during the exploration phase, which demonstrates the effectiveness of the optimizer in exploring the search space.
Furthermore, in the proximity of the optimal solution which is represented by the yellow star, there is a noticeable convergence trend of particle locations towards the optimum location. 
This observation suggests that efficient exploitation steps have been performed to obtain the optimum solution during the optimization process.

%%%%%%%%%%%%%BEGIN %FIGURE%%%%%%%%%%%%%%%%
\begin{figure}
\centering
\includegraphics[width=.65\linewidth]{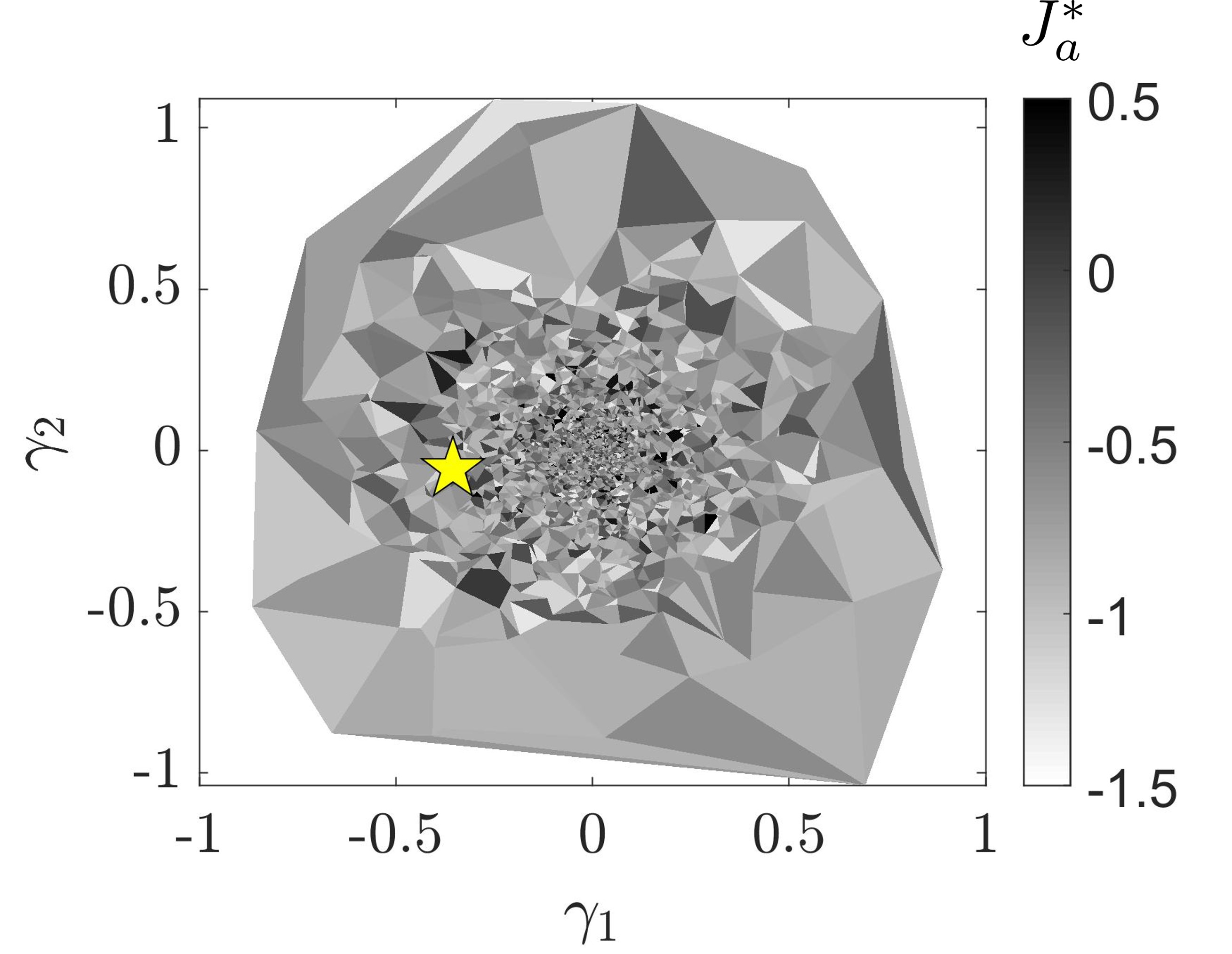}
\caption{The two-dimensional proximity map from PSO-TPME optimization.}
\label{fig:proximity_map}
\end{figure}
%%%%%%%%%%%%%%%END %FIGURE%%%%%%%%%%%%%%%%

We visualize the control landscape topology based on the optimization data in the best run case, and the results are displayed in \Cref{fig:proximity_map}.
In this study, the proximity map is computed using the classical multi-dimensional scaling \citep{Cox2000}, and the location of the optimal parameter is highlighted with a yellow star.
The parameters $\gamma_1$ and $\gamma_2$ represent the two-dimensional compression of the 60-dimensional distances among the particles. 
During the calculation, all input parameters are re-scaled into the range of $[-1,1]$ to ensure equal weighting.
Under the two-dimensional representation, the control landscape topology appears to be extremely rugged. 
There are several potential reasons for this phenomenon. 
Firstly, it could be attributed to the oversimplification of the 60-dimensional feature vector, where projecting a highly non-convex control landscape onto a two-dimensional plane results in a highly non-smooth representation. 
Secondly, the configuration of the actuators in the Smart Skin could contribute to the ruggedness of the landscape. 
Similar to the comparison between \Cref{fig:parametric1} and \Cref{fig:parametric2}, where the inclusion of active control leads to completely different flow control performances compared to the passive-only cases, slight differences in the flow control configuration among neighboring particles in the search space can yield significantly different outcomes.
Overall, the ruggedness of the control landscape topology demonstrates the complexity of the optimization problem for finding effective control parameters in the Smart Skin, and proves the effectiveness of PSO-TPME to tackle complex high-dimensional search space within a finite time horizon.

%%%%%%%%%%%%%BEGIN %FIGURE%%%%%%%%%%%%%%%%
\begin{figure}
\centering
\includegraphics[width=.6\linewidth]{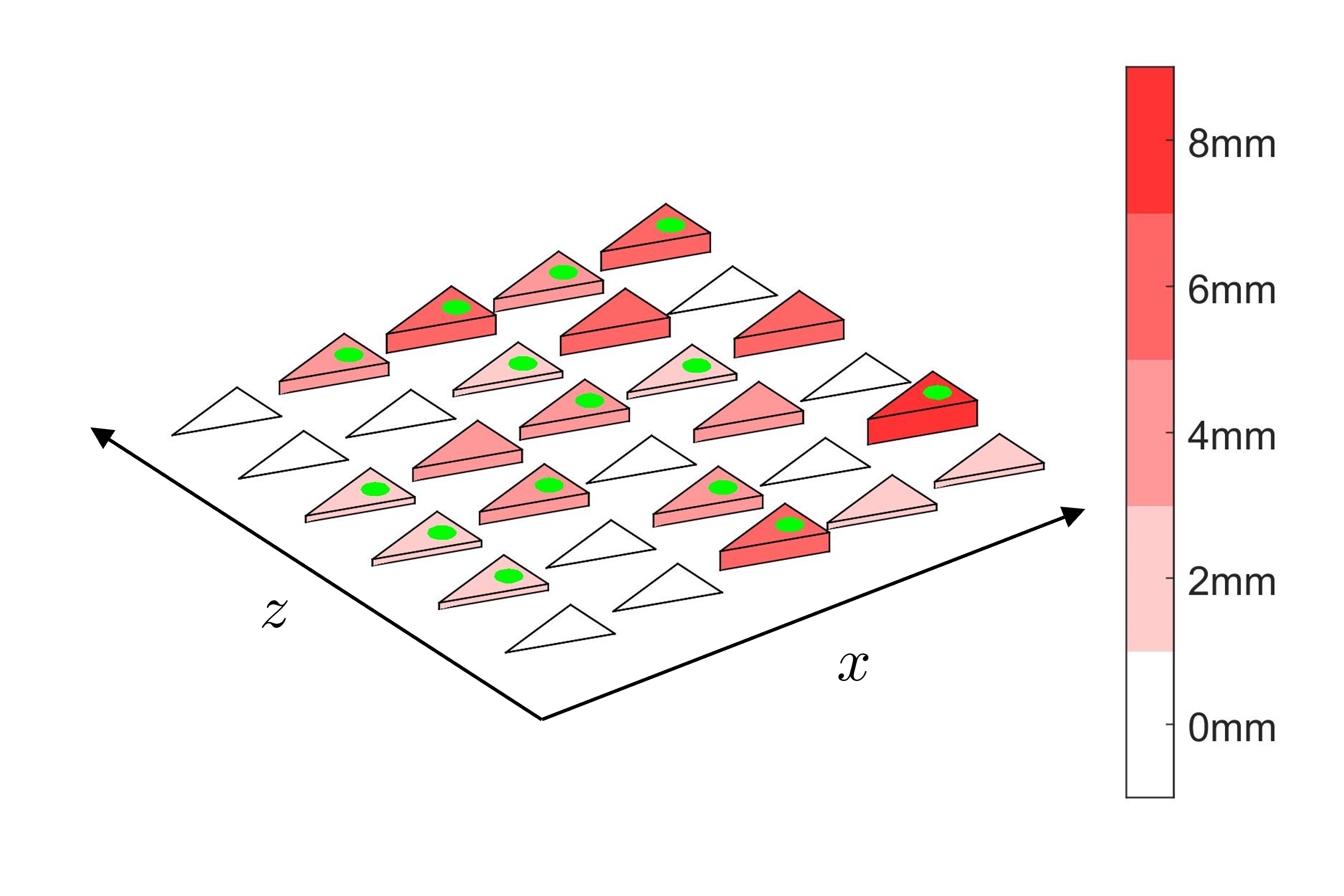}
\caption{Visualization of the best Smart Skin control parameters from PSO-TPME.}
\label{fig:pso_best}
\end{figure}
%%%%%%%%%%%%%%%END %FIGURE%%%%%%%%%%%%%%%%

\Cref{fig:pso_best} provides a visualization of the control parameters from the best individual obtained through the PSO-TPME optimization. 
In this visualization, the passive control is represented by the heights of the corresponding actuators (with color coding), while the active actuations are depicted as green dots.
Despite the actuation pattern not being strictly symmetric about the middle line, valuable insights can still be gained from these actuation parameters. 
Consistent with the findings from the parametric study, the best individual in this optimization exhibits active actuation at the minimum height of 2 mm on the first spanwise row. 
Additionally, active blowing at the third row is also a significant feature in this actuation pattern.
Regarding passive control, most of the activated actuators maintain a height between 2 and 4 mm. 
Actuators with heights larger than 4 mm are primarily located near the side walls and on the last row.

The asymmetry observed in the optimized actuation pattern can be attributed to several factors. 
Firstly, in this 60-parameter optimization experiment, there is no enforced assumption of symmetry. 
This results in an extremely high-dimensional search space, and it becomes challenging to obtain a strictly symmetric actuation pattern within a limited time budget considering the large search space.
Furthermore, the asymmetry in the actuation pattern may also arise from the presence of actuators that are insignificant in terms of their impact on the flow control outcome. 
If the activation status of certain actuators does not significantly affect the flow control performance, the resulting actuation pattern may not exhibit strong symmetry. 
Analyzing the importance of input parameters, similar to the work conducted in \citet{menze2009}, is a valuable avenue for future research. 
By assessing the significance of individual input parameters and enforcing symmetry, we can effectively identify the key parameters that drive the performance of the Smart Skin. This also allows us to reduce the dimensional complexity of the optimization problem and ultimately achieve improved flow control performance within the same time budget.

%%%%%%%%%%%%%BEGIN %FIGURE%%%%%%%%%%%%%%%%
\begin{figure}
\centering
\includegraphics[width=.6\linewidth]{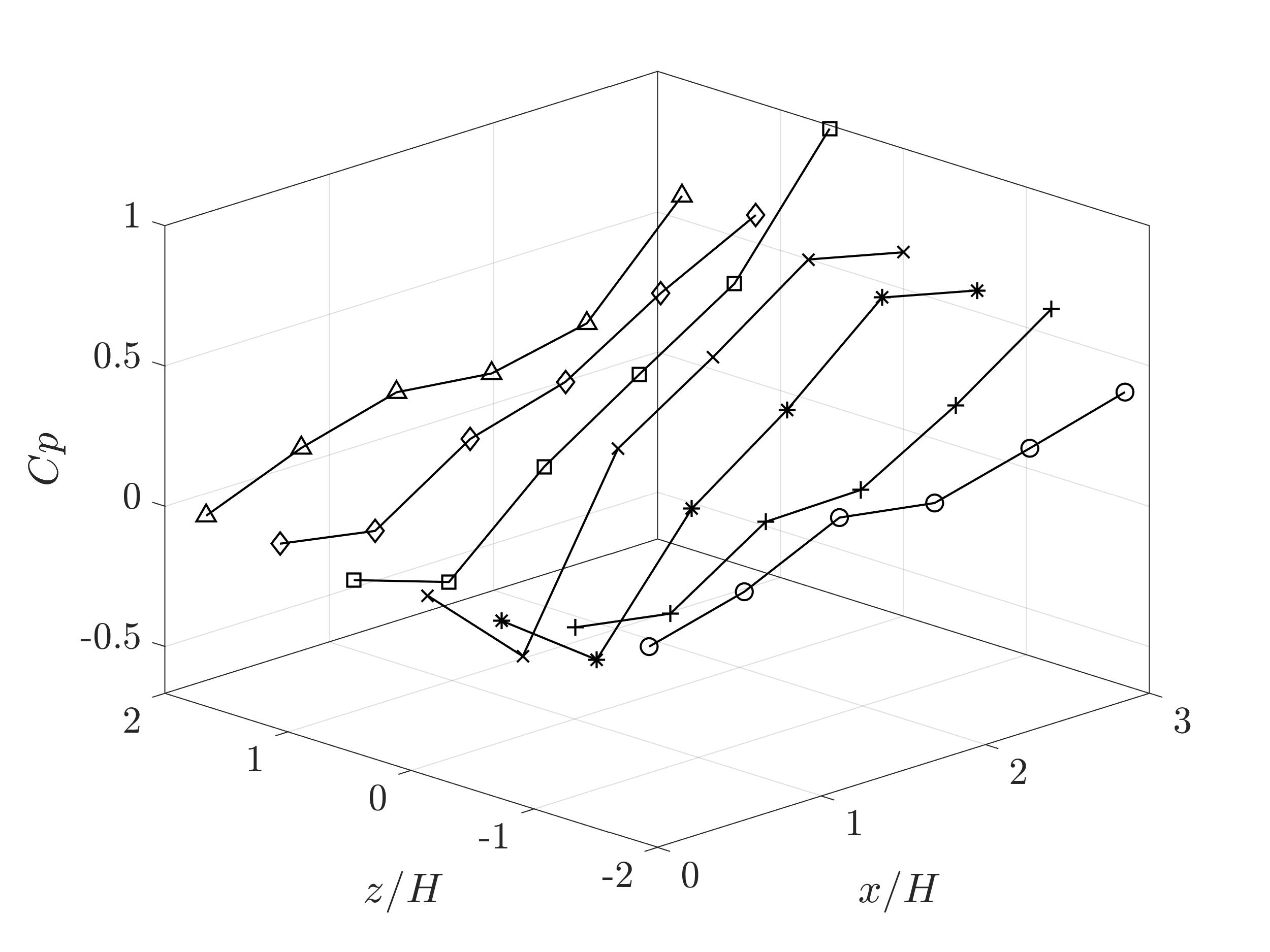}
\caption{The distribution of pressure coefficient $C_p$ on the ramp surface under the optimized control parameters.}
\label{fig:cp_pso}
\end{figure}
%%%%%%%%%%%%%%%END %FIGURE%%%%%%%%%%%%%%%%

The pressure distribution under the optimized flow control parameters is presented in \Cref{fig:cp_pso}.
The first-row blowing in the Smart Skin induces a gradually decrement of $Cp$ in the streamwise direction.
After experiencing the sudden pressure drop, the pressure on the ramp surface begins to gradually recover in the streamwise direction. 
The introduction of downstream blowing appears to contribute to an additional increase in pressure recovery on the aft part of the ramp surface. 
As a result, the pressure coefficient rises above 0.5 at a downstream distance of $3H$.

%%%%%%%%%%%%%BEGIN %FIGURE%%%%%%%%%%%%%%%%
\begin{figure}
\centering
\includegraphics[width=.7\linewidth]{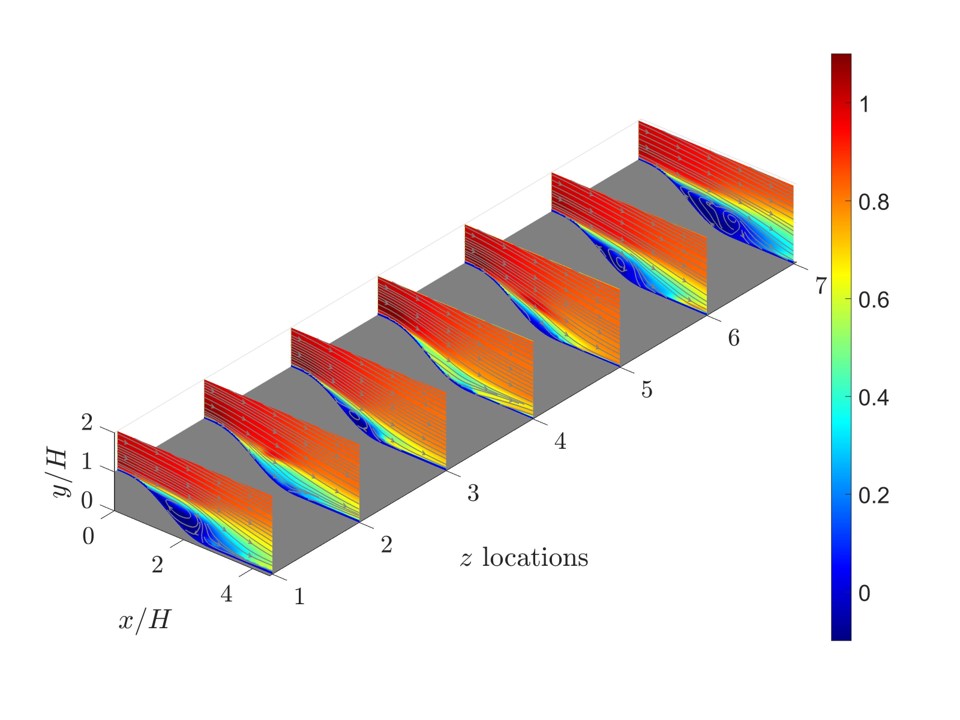}
\caption{Streamwise velocity contours and the superpositioned 2D streamlines measured at 7 spanwise locations with PIV under the optimized control parameters. The spanwise locations are aligned with the locations of the pressure taps. The $z$ locations 1 to 7 correspond to $z/D=$ - 1.8, 1.2, 0.6, 0, 0.6, 1.2, and 1.8, respectively.}
\label{fig:PIV_pso_1}
\end{figure}
%%%%%%%%%%%%%%%END %FIGURE%%%%%%%%%%%%%%%%

%%%%%%%%%%%%%BEGIN %FIGURE%%%%%%%%%%%%%%%%
\begin{figure}
\centering
\includegraphics[width=.8\linewidth,trim={0cm 1.5cm 0cm 1.5cm},clip]{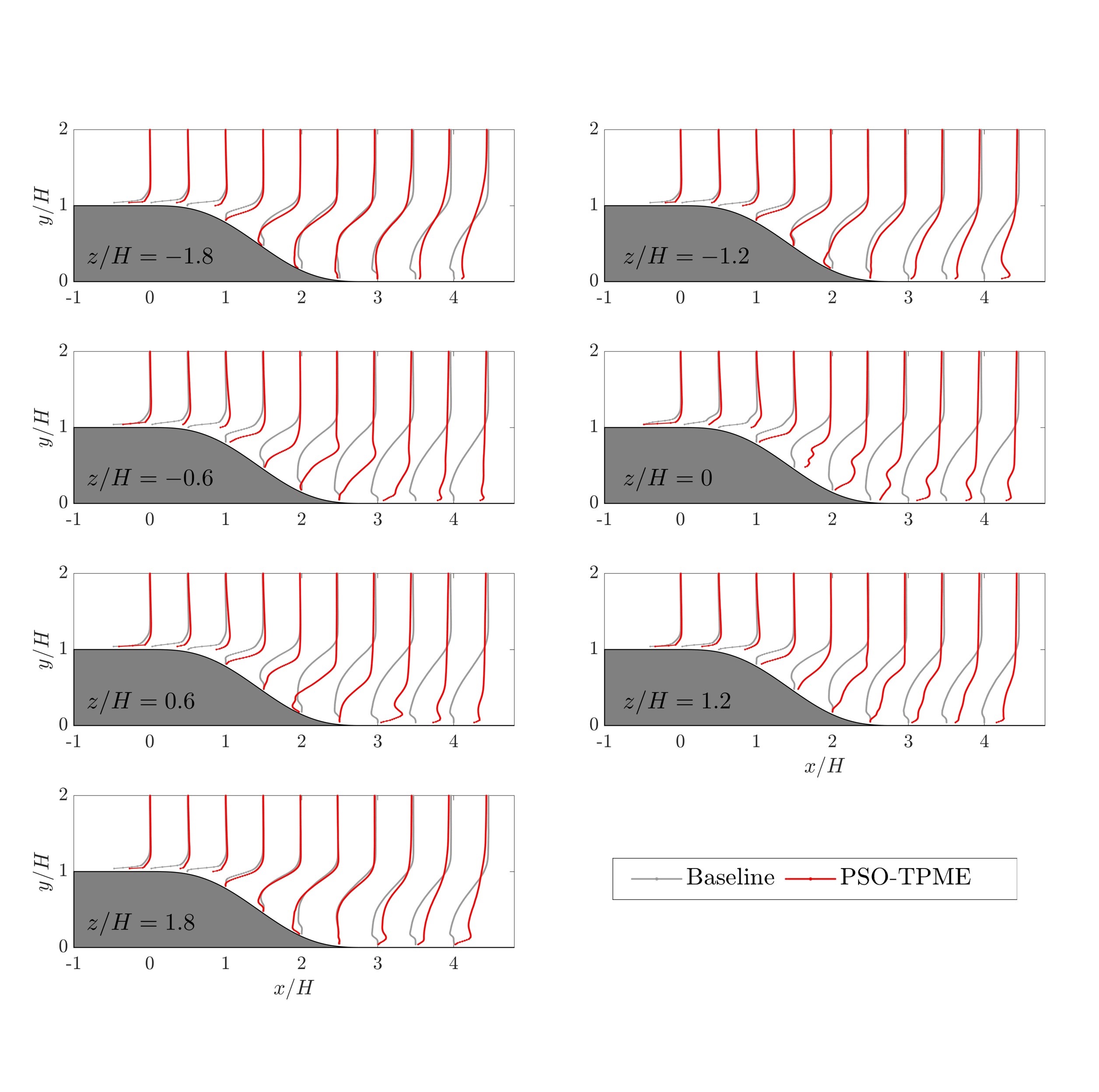}
\caption{Comparison of streamwise velocity profiles at different spanwise locations under the baseline and PSO-TPME optimization cases.}
\label{fig:PIV_pso_2}
\end{figure}
%%%%%%%%%%%%%%%END %FIGURE%%%%%%%%%%%%%%%%

\Cref{fig:PIV_pso_1} visualize the velocity profiles obtained from the PIV measurements under the optimum control law.
A comparison of the flow profiles from the baseline and the optimal controlled cases is also provided in \Cref{fig:PIV_pso_2}. 
Significant separation reduction can be obtained from the optimized control strategy where the reduced area of the low-speed region can be observed at nearly all cases except for the locations near the side wall.
The performance at these near-wall locations is influenced by the actuation from only one side of the measurement plane, yet a reduced recirculation area in the downstream location can be already observed compared to the unforced case.
The improved flow field demonstrates the success of flow control optimization using PSO-TPME, which is capable to conquer the high-dimensional search space within a finite time budget.

\section{Conclusions and outlook}\label{sec:S4}

In this work we design and construct the Smart Skin, a separation control device featuring multi-modal actuators and distributed-input, distributed-output (DIDO).
Each multi-modal actuator can execute quasi-static passive control --- similar to vortex generators, and can implement additional active control using minijet actuation.
We deploy these actuators on the surface of a ramp in a distributed manner, and also deploy distributed pressure taps among the actuators to monitor the flow state.
This DIDO system can provide open- and closed-loop control based on the different configurations of the control algorithms.

We characterize the unforced flow and perform a parametric study of the Smart Skin.
We found the unforced flow follows the characteristics of the flow around a smooth ramp.
The dominant turbulent event is the flapping motion of the recirculation region in the wall-normal direction, followed by the convection of coherent structures in the shear layer.
Among the parametric cases, the best passive control occurs when the second and the third spanwise rows are lifted up to \SI{8}{\milli\meter}, and the best active + passive control is achieved when the first two rows blow at the minimum height of \SI{2}{\milli\meter}.
Furthermore, the flow under the best passive control strategy is not further improved after adding active control in the flow.
These observations demonstrate a very complex control landscape, and necessitate the need for machine learning optimization of the control laws.

Metaheuristic optimization algorithms based on machine learning represents a set of fast-converging optimization algorithms which can conquer the high-dimensional search space.
In this work we utilize the cutting-edge variant of the particle swarm optimization \citep{Shaqarin2023} to optimize the open-loop control parameters in the Smart Skin.
This forms a 60-dimensional optimization problem under a finite time horizon.
Yet, PSO-TPME can effectively tackle the high-dimensional problem and leads to pressure recovery which is much better than the parametric result.
Although the optimized control law is not fully symmetric,
by adding prior symmetrical constraints and identifying non-significant parameters, the efficiency of PSO-TPME can be further improved.

The successful story of the Smart Skin demonstrates the feasibility of deploying DIDO flow control devices on a wide range of flow control plants.
For example, we can naturally extend the current configuration to the slanted surface of an Ahmed body.
Another example would be the distributed deployment of multi-modal actuators on an airfoil for separation control.
For industrial applications, flow control in transport vehicles can also benefit from similar configurations with distributed sensors and actuators.

In addition, adding sensor feedback may further enhance flow separation control \citep{Brunton2015}.
For model-based control, sensor signals can be utilized to estimate flow state.
Classical examples can be found in \citet{Samimy2007,Bieker2020}, where the quadratic stochastic estimation and deep model predictive control are integrated into the model-based controllers.
For model-free control, the sensor signals can also provide essential flow information to feedback functions.
Machine learning control (MLC,  \citealt{Duriez2017,CornejoMaceda2021,xmlc2022}) and reinforcement learning (RL, \citealt{Verma2018,Rabault2019,Fan2020,Vinuesa2022}) have demonstrated huge potential for model-free control with sensor signals.
The involvement of sensor signals in the feedback flow control can potentially improve flow control performance, and the interpretation of the control laws can also enhance the understanding of the flow control outcome.
These potential extensions of the Smart Skin shed light on the future flow control paradigms with distributed sensors, actuators, and machine learning.

\section*{Funding}
This work is supported by the China Aerospace Science and Technology Corporation (CASC) Innovative Research Project XFX20220401.
\hl{The author acknowledges Prof. Bernd R. Noack for his selfless help on this work.}

\section*{Declaration of interests}
The authors report no conflict of interest.

\section*{Author ORCIDs}

Songqi Li, \url{https://orcid.org/0000-0002-8737-8334}; \\
Tamir Shaqarin \url{https://orcid.org/0000-0003-3327-3147};\\
Bernd R. Noack, \url{https://orcid.org/0000-0001-5935-1962}.

\bibliographystyle{jfm}
\bibliography{references}

%% End of file `jfm2esam.bib'.

\end{document}